\documentclass[pra,aps,twocolumn,amsfonts,showpacs,superscriptaddress]{revtex4}

\usepackage{amsmath}
\usepackage{amsfonts}
\usepackage{amssymb}
\usepackage{graphicx}
\usepackage{epstopdf}
\usepackage{epsfig}
\usepackage{psfrag}
\usepackage{color}
\usepackage{subfigure}
\usepackage{xspace}

\newcommand{\ket}[1]{|#1\rangle}

\newcommand{\om}[0]{\omega}
\newcommand{\en}[0]{\epsilon}

\newcommand{\las}[0]{\langle}
\newcommand{\ras}[0]{\rangle}

\newcommand{\nag}{{\phantom{\dag}}}
\newcommand{\rmi}{\mathrm{i}}

\begin{document}

\title{Bose--Hubbard Models Coupled to Cavity Light Fields} 

\author{A. O. Silver}
\affiliation{University of Cambridge, Cavendish Laboratory, Cambridge,
CB3 0HE, UK.} 
\author{M. Hohenadler}
\affiliation{University of Cambridge, Cavendish Laboratory, Cambridge,
CB3 0HE, UK.}  
\affiliation{OSRAM Opto Semiconductors
GmbH, 93055 Regensburg, GER.}  
\author{M. J. Bhaseen} 
\affiliation{University of Cambridge, Cavendish
Laboratory, Cambridge, CB3 0HE, UK.}  
\author{B. D. Simons} 
\affiliation{University of
Cambridge, Cavendish Laboratory, Cambridge, CB3 0HE, UK.}
\date{\today}

\begin{abstract}
  Recent experiments on strongly coupled cavity quantum
  electrodynamics present new directions in ``matter--light'' systems.
  Following on from our previous work [Phys. Rev. Lett. {\bf 102},
  135301 (2009)] we investigate Bose--Hubbard models coupled to a
  cavity light field.  We discuss the emergence of photo-excitations
  or ``polaritons" within the Mott phase, and obtain the complete
  variational phase diagram.  Exploiting connections to the
  superradiance transition in the Dicke model we discuss the nature of
  polariton condensation within this novel state.  Incorporating the
  effects of carrier superfluidity, we identify a first order
  transition between the superradiant Mott phase and the single
  component atomic superfluid. The overall predictions of mean field
  theory are in excellent agreement with exact diagonalization and we
  provide details of superfluid fractions, density fluctuations, and
  finite size effects. We highlight connections to recent work on
  coupled cavity arrays.
\end{abstract}


\maketitle

\section{Introduction}
Over the last few years there has been tremendous activity in studying
the coherent interaction of matter and radiation in a rich variety of
``matter--light'' systems. Recent experiments have combined cavity
quantum electrodynamics (cavity QED) with cold atomic gases, and allow
access to the strongly coupled regime.  This has led to pioneering
work on Bose--Einstein condensates (BECs) in ultra high finesse
optical cavities \cite{Brennecke:Cavity}, and with optical fibers on
atom chips \cite{Colombe:Strong}. It has also stimulated advances in
cavity opto-mechanics and condensate dynamics
\cite{Brennecke:Opto,Ritter:Dyn}.  More recently, strong matter--light
coupling has been achieved for ion crystals
\cite{Herskind:Ion,Lange:Strength}, with potential applications in
quantum information processing.  These developments offer a wealth of
possibilities, at the interface between quantum optics, cold atoms and
condensed matter physics. The light field serves not only as a probe
of the many-body system \cite{Ritsch:Probing}, but may also support
interesting cavity mediated phenomena and phases. It may further
provide routes to simulate strongly correlated quantum systems, with
proposals based on coupled cavity arrays
\cite{Hartmann:Strongly,Greentree:QPTlight,Angelakis:Blockade,Na:Polarray,Hartmann:Rev,Cho:Fractional,Illuminati:Light,Rossini:JCH,Lei:QPT,Makin:QPT,Aichhorn:JCH,Pippan:Spectra,Groshol:Nanocavity,Schmidt:Strong,Mering:JCH,Koch:JCL}
and nonlinear optical fibers \cite{Chang:Cryst}.

Allied advances in solid state devices include cavity QED experiments
with superconducting qubits in microwave resonators
\cite{Wallraff:Strong}. This has provided clean realizations of the
paradigmatic Jaynes--Cummings \cite{Fink:Climbing} and Dicke models
\cite{Wallraff:Collective}, describing two-level systems coupled to
radiation. It has also led to remarkable observations of the Lamb
shift \cite{Wallraff:Lamb}. This is complemented by the quest for
polariton condensates in semiconductor microcavities
\cite{Kasprzak:BEC,Deng:Cond,Deng:Lasing,Dang:Stim,Littlewood:Models,Kavokin:Micro},
where the hybridization of an exciton and a photon yields low
effective mass polaritons.  This offers the prospect of higher
transition temperatures than for exciton BEC, and gives access to
coherence properties via the cavity light field.

Motivated by this broad spectrum of activity, we examine the impact of
cavity radiation on bosonic Hubbard models
\cite{Zoubi:Expol}. Following on from our previous work
\cite{Bhaseen:Polaritons}, we focus on a two-band model in which
photons induce transitions between two internal states or Bloch
bands. This is a natural generalization of the much studied two-level
systems coupled to radiation, and may serve as a useful paradigm in
other contexts.  In Ref.~\cite{Bhaseen:Polaritons} we discussed the
interplay of Mott physics, photo-excitation, and Bose condensation
promoted by carrier itinerancy. In particular, we provided evidence
for a novel Mott phase with photo-excitations analogous to
polaritons. In this work we study this problem in more detail, with
emphasis on the nature of the polariton condensate.  We also highlight
connections to coupled cavity arrays described by the
Jaynes--Cummings--Hubbard model and its variants
\cite{Hartmann:Strongly,Greentree:QPTlight,Angelakis:Blockade,Na:Polarray,Hartmann:Rev,Cho:Fractional,Groshol:Nanocavity,Illuminati:Light,Rossini:JCH,Lei:QPT,Makin:QPT,Aichhorn:JCH,Schmidt:Strong,Mering:JCH,Koch:JCL}. Additional
directions in cold atoms include recent work on excitons
\cite{Kantian:ALE}, generalized Dicke models \cite{Larson:Dilute}, and
light propagation in atomic Mott insulators
\cite{Bariani:Prop,Carusotto:Slow}.

The outline of this paper is as follows. We begin in section
\ref{Sect:Model} with an introduction to the two-band Bose--Hubbard
model coupled to quantum light \cite{Bhaseen:Polaritons}. In section
\ref{Sect:Zero} we discuss the zero hopping limit of this model and
anchor the phase diagram.  We study the evolution with the strength of
the matter--light coupling and highlight the connections to the Dicke
model and the superradiance transition
\cite{Dicke:Coherence,Hepp:Super,Wang:Dicke,Hepp:Eqbm}. In section
\ref{Sect:Hopping} we use a variational approach to obtain the overall
phase diagram in the presence of hopping.  We corroborate our findings
in section \ref{Sect:Numerics} using numerical simulations.  We
discuss the relation to the bosonic BEC--BCS crossover and to other
problems of current interest in section \ref{Sect:Other}.  We conclude
in section \ref{Sect:Conc} and provide technical appendices.

\section{Model}
\label{Sect:Model}
\begin{figure}
\begin{center}
  \includegraphics[width=6cm,clip=true]{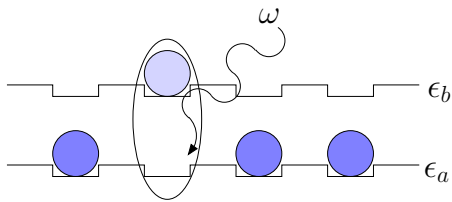}
\caption{(Color online). Heuristic diagram showing photo-excitations above the 
filled bosonic Mott state. In analogy with fermionic band insulators,
the coherent superposition of a particle--hole pair and a photon
will be referred to as a ``polariton''. 
Within an equilibrium framework these polaritons may Bose condense.
\label{Fig:bhpolariton}}
\end{center}
\end{figure}

Let us consider a two-band Bose--Hubbard model coupled to the quantum
light field of an optical cavity in the rotating wave approximation
\cite{Bhaseen:Polaritons}:
\begin{equation}
\begin{aligned}
H_0 & = \sum_{i\alpha} \epsilon_\alpha n_{i\alpha}+\omega
\psi^\dagger\psi + \sum_{i\alpha\alpha^\prime}
\frac{U_{\alpha\alpha^\prime}}{2}:n_{i\alpha}n_{i\alpha^\prime}: \\ &
-\sum_{\langle ij\rangle} J_\alpha
(\alpha_i^\dagger\alpha_j^{\phantom\dagger}+{\rm h.c.})+ g\sum_i
(b_i^\dagger a_i^{\phantom\dagger}\psi+{\rm h.c.}),
\end{aligned}
\label{qlmodel}
\end{equation}
where $i$ labels lattice sites, and $\alpha=a,b$ are bosons obeying
canonical commutation relations
$[\alpha_i^{\phantom\dagger},\alpha_j^\dagger]=\delta_{ij}$. These
might be states of different orbital or spin angular momentum, and we
assume the cavity radiation field, $\psi$, may induce
photo-excitations between these states; see
Fig.~\ref{Fig:bhpolariton}.
Here, $\epsilon_\alpha$ effects the band splitting $\omega_0\equiv
\epsilon_b-\epsilon_a$, and $U_{\alpha\alpha^\prime}$ are
interactions, where $::$ indicates normal ordering.  This yields
$:n_{i\alpha}n_{i\alpha}:=n_{i\alpha}(n_{i\alpha}-1)$ for like species
and $:n_{i\alpha}n_{i\alpha^\prime}:=n_{i\alpha}n_{i\alpha^\prime}$
for distinct species.  $J_\alpha$ are nearest neighbor hopping
parameters, and $\omega$ is the frequency of the cavity mode. For
simplicity we consider just a single mode which couples uniformly to
the bands.  The coupling $g$ is the strength of the matter--light
interaction.  In view of the box normalization of the photon field,
the dipole coupling strength is proportional to $1/\sqrt{V}$, where
$V$ is the volume of the cavity. With a fixed density of lattice
sites, $\rho=N/V$, it is convenient to denote $g\equiv \bar
g/\sqrt{N}$, where $N$ is the total number of lattice sites. We work
in units where the half-band splitting
$\omega_0/2=(\epsilon_b-\epsilon_a)/2=1$.

An important feature of the Hamiltonian (\ref{qlmodel}) is that the
individual atom and photon numbers are {\em not} conserved due to the
matter--light interaction. However, the total number of atomic
carriers
\begin{equation}
N_1\equiv \sum_i (n_i^b+n_i^a),
\end{equation}
and the total number of photo-excitations 
\begin{equation}
N_2\equiv \psi^\dagger\psi+\frac{1}{2}\sum_i(n_i^b-n_i^a+1),
\end{equation} 
are conserved and commute with $H_0$. The latter counts the total
number of photons plus particle--hole pairs, and we refer to these
composite excitations as ``polaritons'' --- see
Fig.~\ref{Fig:bhpolariton}.  These conservation laws reflect the
global ${\rm U}(1)\times {\rm U}(1)$ symmetry of $H_0$ such that
\begin{equation}
a\rightarrow e^{i\vartheta}a,\quad b\rightarrow e^{i\varphi}b, \quad 
\psi\rightarrow e^{-i(\vartheta-\varphi)}\psi,
\end{equation}
where $\vartheta,\varphi\in{\mathbb R}$. In general this symmetry
involves mixing between the matter--light sectors. Moreover, it also
allows for the simultaneous coexistence of Mott behavior and
condensation, corresponding to an unbroken ${\rm U}(1)$ and broken
${\rm U}(1)$ symmetry respectively. This symmetry will have a direct
manifestation in the phase diagram, and will suggest implications for
other multicomponent problems.  We will work in the grand canonical
ensemble:
\begin{equation}
H=H_0 -\mu_1 N_1-\mu_2 N_2.
\label{gham}
\end{equation}
We begin by assuming that $a$ are strongly interacting hardcore
bosons, $U_{aa}\rightarrow\infty$, and that $b$ are sufficiently
dilute so that we may neglect their interactions, $U_{bb}=0$. We will
also start with $U_{ab}=0$, before discussing departures from these
conditions.

\section{Zero Hopping Limit}
\label{Sect:Zero}
Before embarking on a detailed examination of the model
(\ref{qlmodel}), we investigate the zero hopping limit.  As in the
single-band Bose--Hubbard model \cite{Fisher:Bosonloc,Greiner:SI} this
anchors the topology of the general phase diagram. In the present case
this is particularly informative since the zero hopping phase diagram
evolves with the matter--light coupling, $g$. In addition, the global
photon mode couples {\em all} the sites, even in this zero hopping
limit.  To gain a handle on this reduced Hamiltonian, we find it
convenient to develop two complementary approaches. These help
illuminate different aspects of the more general itinerant problem,
and provide a platform for extensions. In section \ref{Sect:Zerovar}
we begin with a variational approach using a coherent state ansatz for
the photons. This will enable us to derive an effective single site
Jaynes--Cummings model \cite{Jaynes:Comp} for the $a,b$ bosons, and
proceed with the minimum of technical input. In section
\ref{Sect:Dicke} we instead map the $a,b$ bosons on to effective
spins. This yields the paradigmatic Dicke model
\cite{Dicke:Coherence}, describing many spins coupled to
radiation. The Jaynes--Cummings and Dicke models are familiar in
atomic physics and quantum optics and we provide a brief overview of
these closely related Hamiltonians in Appendix \ref{App:Dicke}. Both
approaches yield equivalent results and indicate a novel quantum phase
transition occurring within the lowest Mott lobe
\cite{Bhaseen:Polaritons}. This takes place from a conventional Mott
state with no photons, to a state with a non-vanishing population of
photo-excitations, as the strength of the matter--light coupling is
increased. This quantum phase transition coincides with the well known
superradiance transition in the Dicke model
\cite{Dicke:Coherence,Hepp:Super,Wang:Dicke,Hepp:Eqbm}, and we
interpret the results in this framework.  The combination of
perspectives will be useful in the generalization to the itinerant
problem discussed in section \ref{Sect:Hopping}. For other proposals
of Dicke models and superradiance transitions in cold atomic gases see
Refs.~\cite{Dimer:Proposed,Chen:Exotic,Zoubi:SR}.

\subsection{Variational Approach}
\label{Sect:Zerovar}
In the zero hopping and $U_{aa}\rightarrow \infty$ limit, the
Hamiltonian (\ref{gham}) may be written in the form
\begin{equation}
\begin{aligned}
H & =\tilde\epsilon_a\sum_i n_i^a+\tilde\epsilon_b \sum_i
n_i^b+\tilde\omega\psi^\dagger\psi\\ & + g\sum_i(b_i^\dagger
a_i^{\phantom\dagger}\psi+\psi^\dagger a_i^\dagger
b_i^{\phantom\dagger}),
\label{reduced}
\end{aligned}
\end{equation}
where the site occupancy of $a$-atoms is limited to $0,1$. We absorb
the chemical potentials in to the definitions
\begin{equation}
\begin{aligned}
\tilde\epsilon_a & \equiv \epsilon_a-\mu_1+\mu_2/2,\\
\tilde\epsilon_b & \equiv \epsilon_b-\mu_1-\mu_2/2,\\
\tilde\omega & \equiv \omega-\mu_2.
\end{aligned}
\end{equation}
This bosonic Hamiltonian is analogous to the fermionic problem of
localized excitons in a microcavity
\cite{Eastham:Localized,Eastham:Beyondlin}. An instructive way to
analyze the problem is to consider a coherent state for the cavity
light field:
\begin{equation}
|\gamma\rangle\equiv e^{-\frac{\gamma^2}{2}+\gamma\psi^\dagger}|0\rangle,
\label{cs}
\end{equation}
where $\gamma$ is a variational parameter to be determined. As usual,
this is an eigenstate of the annihilation operator, $\psi$, with
eigenvalue $\gamma$. The coherent state assumption is expected to
become exact in the thermodynamic limit, and this is borne out by the
complementary approach in section \ref{Sect:Dicke}.  The expectation
value
\begin{equation}
\langle \gamma |H|\gamma\rangle\equiv \sum_i H_{\rm eff},
\end{equation}
yields an effective {\em single site} problem for the $a$ and $b$ atoms
which is readily diagonalized:
\begin{equation}
H_{\rm eff}=\tilde\epsilon_an^a+\tilde\epsilon_bn^b+
\bar\omega\gamma^2+g\gamma(b^\dagger a+a^\dagger b),
\label{heff}
\end{equation}
where $\bar\omega\equiv\tilde\omega/N$. This is a significant merit of
the current approach, since it is directly tractable without technical
input.  In view of the hardcore constraint on the $a$ atoms, this
describes a {\em single} two-level system coupled to an effective
``radiation field'' of $b$ atoms. This paradigm is described by the
much studied Jaynes--Cummings model \cite{Jaynes:Comp,Cummings:Stim},
discussed in Appendix \ref{App:Dicke}. The eigenstates of (\ref{heff})
are superpositions in the upper and lower bands. Focusing on Mott
states with total occupation $n_a+n_b=n$, we consider admixtures of
$|0,n\rangle$ and $|1,n-1\rangle$ in the $|n_a,n_b\rangle$ basis. The
lowest eigenstate has energy
\begin{equation}
E_n^-=\bar\omega\gamma^2+n\tilde\epsilon_b-\tilde\omega_0/2
-\sqrt{\tilde\omega_0^2/4+g^2\gamma^2 n},
\label{enminus}
\end{equation}
where $\tilde\omega_0\equiv \tilde\epsilon_b-\tilde\epsilon_a$ is the
effective band splitting. The nonlinear dependence on $\sqrt{n}$ is a
notable feature of the Jaynes--Cummings eigenstates and has recently
been seen in circuit QED experiments \cite{Fink:Climbing}.  (Analogous
dependence is also seen in BECs \cite{Brennecke:Cavity,Colombe:Strong}
as a function of the number of atoms in the cavity.)  Minimizing with
respect to $\gamma$, one obtains the variational self-consistency
condition
\begin{equation}
\frac{\partial
E_n^-}{\partial\gamma}=2\gamma\left(\bar\omega-\frac{g^2
n}{\sqrt{\tilde\omega_0^2+4g^2\gamma^2 n}}\right)=0.
\label{selfcon}
\end{equation}
Depending on the parameters, one may therefore obtain either the
trivial solution $\gamma=0$, corresponding to zero photon occupancy,
or the non-trivial solution
\begin{equation}
\gamma^2_{\rm var}=\frac{1}{4}\left(\frac{g^2n}{\bar\omega^2}
-\frac{\tilde\omega_0^2}{g^2n}\right)=\langle\psi^\dagger\psi\rangle,
\label{gammavar}
\end{equation}
corresponding to a finite photon occupancy. This latter solution is
supported in the region where $\gamma_{\rm var}^2>0$, or when the
matter--light coupling exceeds the critical value
\begin{equation}
\bar {\mathcal G}_c\equiv \bar g_c\sqrt{n}=\sqrt{\tilde\omega\tilde\omega_0},
\label{gc}
\end{equation}
where for simplicity we consider $\tilde\omega_0>0$. 
As we shall discuss in section \ref{Sect:Dicke}, this onset of the
photon field corresponds to the well known superradiance transition in
the Dicke model
\cite{Dicke:Coherence,Hepp:Super,Wang:Dicke,Hepp:Eqbm}. Indeed, it is
readily seen from equation (\ref{gammavar}), that for $\bar {\mathcal
  G}\ge \bar {\mathcal G}_c$, the photon field has the characteristic
variation
\begin{equation}
\frac{\langle \psi^\dagger
\psi\rangle}{N}=\frac{1}{4\tilde\omega^2}\left(\frac{\bar {\mathcal
G}^4-\bar {\mathcal G}_c^4}{\bar {\mathcal G}^2}\right),
\label{photonset}
\end{equation}
where $\bar{\mathcal G}={\bar g}\sqrt{n}$. See for example Table 1 of
Ref.~\cite{Emary:Chaos}. We shall give an alternative derivation of
these results in section \ref{Sect:Dicke}.  This complementary
approach becomes asymptotically exact in the thermodynamic limit and
helps justify the coherent state ansatz (\ref{cs}).

\begin{figure}
\includegraphics[width=8cm,clip=true]{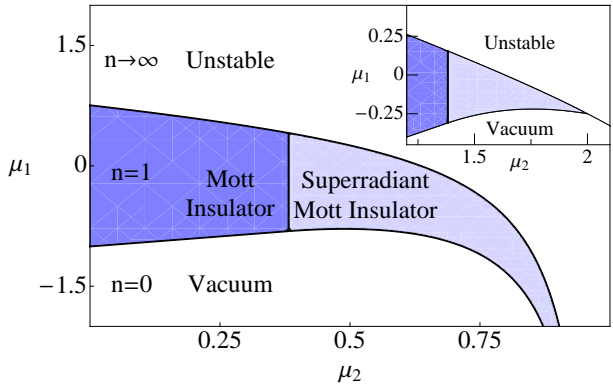}
\caption{(Color online). Zero hopping phase diagram in the
  $U_{aa}\rightarrow\infty$ limit with $U_{bb}=U_{ab}=0$. We set
  $\epsilon_b=-\epsilon_a=\bar g=1$ and $\omega=1$ corresponding to
  $\omega<\omega_0$, where $\omega_0\equiv \epsilon_b-\epsilon_a$. The
  vertical line, $\bar g=\bar g_c$, is the superradiance transition in
  the Dicke model, and separates a Mott insulator (MI) with
  $n_a+n_b=1$ and $\langle \psi^\dagger\psi\rangle=0$, from a
  superradiant Mott insulator (SRMI) with
  $\langle\psi^\dagger\psi\rangle\neq 0$. Outside of these regions are
  the vacuum state, and the unstable region corresponding to
  macroscopic population of the $b$ states. Whilst the total density
  is fixed within both Mott phases, $n_a+n_b=1$, the individual $a$
  and $b$ populations vary in the superradiant phase as shown in
  Fig.~\ref{Fig:Magonset}. Inset: $\epsilon_b=-\epsilon_a=\bar g=1$
  and $\omega=3$. For $\omega>\omega_0$ the upper and lower boundaries
  may cross and terminate the lobe, since for $\mu_2>\omega_0$ the
  lowest stable state is the vacuum; see Appendix \ref{App:Absence}
  for a derivation of this.}
\label{Fig:zerohop}
\end{figure}

Knowledge of the eigenvalues (\ref{enminus}), together with the
variational consistency condition (\ref{selfcon}), enables one to
construct the zero hopping phase diagram depicted in
Fig.~\ref{Fig:zerohop}.
The lower boundary consists of two segments. The first corresponds to
the transition from the vacuum to the Mott state with no photons, and
is given by the condition $E_1^-(\gamma=0)\le 0$ or
$\mu_1\ge\epsilon_a+\mu_2/2$.  The second segment corresponds to the
transition in to the photon rich Mott state and occurs when
$E_1^-(\gamma_{\rm var})\le 0$ where
\begin{equation}
E_n^-(\gamma_{\rm var})=n\tilde\epsilon_b-\frac{\tilde\omega_0}{2}
-\frac{n}{4\tilde\omega}\left(\frac{\bar
g^4+\bar g_c^4}{\bar g^2}\right).
\end{equation}
Explicitly this locus is given by
\begin{equation}
\mu_1\ge \frac{\epsilon_a+\epsilon_b}{2}-\frac{1}{4\tilde\omega}
\left(\frac{\bar g^4+\tilde\omega^2(\omega_0-\mu_2)^2}{\bar g^2}\right).
\end{equation}
Noting that $\bar g_c$ depends on $n$ according to (\ref{gc}) we see
that $E_n^-(\gamma_{\rm var})\rightarrow n(\tilde\epsilon_b-\bar
g^2/4\tilde\omega)$ as $n\rightarrow\infty$. For
$\tilde\epsilon_b-\bar g^2/4\tilde\omega\le 0$, or chemical potentials
satisfying
\begin{equation}
\mu_1\ge \epsilon_b-\frac{\mu_2}{2}-\frac{\bar g^2}{4(\omega-\mu_2)},
\label{instab}
\end{equation}
it is energetically favorable to macroscopically populate the $b$
levels.  This corresponds to the upper boundary in
Fig.~\ref{Fig:zerohop}. Indeed, in the absence of the matter--light
coupling equation (\ref{instab}) yields $\tilde\epsilon_b\le 0$ and is
naturally associated with population of the $b$ states; we provide the
corresponding $g=0$ phase diagram for comparison in Appendix
\ref{Sect:Absence}. In contrast to the classical light case (where
$\psi$ is treated as a fixed c-number) coupling to the fluctuating
quantum field, $\psi$, eliminates the higher Mott lobes corresponding
to integer increases in the $b$-populations; the higher lobes
corresponding to the $a$-particles have been explicitly eliminated by
the hardcore constraint. We confirm this reduction analytically in
Appendix \ref{App:Absence}.

\begin{figure}
\begin{center}
\includegraphics[width=8cm]{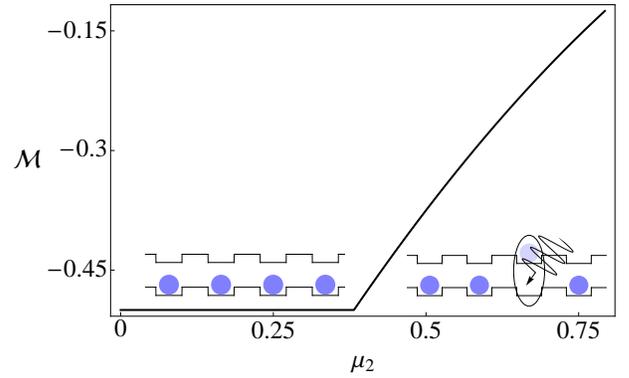}
\caption{(Color online). Onset of magnetization ${\mathcal
    M}=(n_b-n_a)/2$ at the superradiance transition, traversing a line
  through Fig.~\ref{Fig:zerohop} with $\mu_1=0$. Whilst the total
  density $n_a+n_b=1$ remains pinned throughout the entire Mott
  region, photo-excitation promotes atoms between the bands in the
  superradiant phase. \label{Fig:Magonset}}
\end{center}
\end{figure}

Accompanying the superradiance transition is a change in the relative
atomic populations or ``magnetization''
\begin{equation}
\mathcal{M} = \frac{1}{2}\left(\langle n_b\rangle-\langle n_a\rangle\right),
\label{magdef}
\end{equation}
corresponding to photo-excitation into the upper band. Indeed, the
possibility of such magnetic phase transitions is a strong motivation
for studying multicomponent problems, even without matter--light
coupling \cite{Duan:Control,Altman:Twocomp}. The magnetization
(\ref{magdef}) is readily computed from the Jaynes--Cummings
eigenstates (\ref{alphabeta}) with $n=1$. It also serves as an order
parameter for this continuous transition:
\begin{equation}
{\mathcal M}=\begin{cases} 
-\frac{1}{2}; & \bar {\mathcal G}\le \bar {\mathcal G}_c, \\
-\frac{1}{2}\left(\frac{\bar {\mathcal G}_c}{\bar {\mathcal G}}\right)^2; & \bar {\mathcal G}\ge \bar {\mathcal G}_c.
\end{cases}
\label{magev}
\end{equation}
This corresponds to $\langle n_a\rangle=1$ and $\langle n_b\rangle
=0$, for $\bar {\mathcal G}\le \bar {\mathcal G}_c$, and a non-trivial
imbalance, $\langle n_b\rangle -\langle n_a\rangle$, for $\bar
{\mathcal G}>\bar {\mathcal G}_c$.  In Fig.~\ref{Fig:Magonset} we plot
the continuous onset of population imbalance described by equation
(\ref{magev}). In section \ref{Sect:Dicke} we will see how these
results emerge from the exactly solvable Dicke model.

\subsection{Dicke Model}
\label{Sect:Dicke}
As in the fermionic cases considered elsewhere
\cite{Eastham:Localized,Eastham:Beyondlin}, an alternative way to view
the zero hopping Hamiltonian (\ref{reduced}) is as an effective
spin-boson model. Within the subspace of fixed density, $n_a+n_b=n$,
we introduce effective spins for {\em a priori} possible Mott lobes
\begin{equation}
|\Downarrow\rangle = |1,n-1\rangle,\quad |\Uparrow\rangle=|0,n\rangle,
\end{equation}
where we denote bosonic states as $|n_a,n_b\rangle$. The operators
\begin{equation}
S^+=\frac{b^\dagger a}{\sqrt{n}},\quad S^-=\frac{a^\dagger b}{\sqrt{n}},\quad
S^z=\frac{1}{2}-n_a,
\label{hcorespins}
\end{equation}
form a representation of $su(2)$ on this restricted Hilbert space. In
the lowest Mott lobe with $n=1$ this reduces to the usual Schwinger
boson construction \cite{Auerbach:Interacting}.  In the representation
(\ref{hcorespins}) the Hamiltonian (\ref{reduced}) becomes
\begin{equation}
H=\tilde\omega_0\sum_{i=1}^N S_i^z+\tilde\omega\psi^\dagger \psi+
{\mathcal G}\sum_{i=1}^N\left(S_i^+\psi+{\rm h.c.}\right)+c_n,
\label{DHam}
\end{equation}
where ${\mathcal G}\equiv g\sqrt{n}$ and
$c_n=N(n\tilde\epsilon_b-\tilde\omega_0/2)$.  This is the much studied
spin-$1/2$ Dicke model \cite{Dicke:Coherence}, describing $N$
two-level systems coupled to radiation; see Appendix \ref{App:Dicke}
for a brief review. This model is integrable
\cite{Hepp:Super,Hepp:Eqbm,Bog:TC} and in the thermodynamic limit it
has a quantum phase transition to a so-called superradiant phase when
\cite{Hepp:Super,Wang:Dicke,Hepp:Eqbm}
\begin{equation}
\bar{\mathcal G}_c\equiv {\mathcal G}_c\sqrt{N}
=\sqrt{\tilde\omega\tilde\omega_0},
\label{BGC}
\end{equation}
in agreement with our previous result (\ref{gc}). In this context, the
term superradiance indicates the onset of many-body or cooperative
effects involving the photon coupled to many atoms. This mapping not
only helps justify our variational approach, but will also provide
insights into polariton condensation and matter--light coherence at
the superradiance transition.

The thermodynamic limit of the Dicke model (\ref{DHam}) may be
analyzed using collective spin operators \cite{Hillery:Semi}
\begin{equation}
{\bf J}\equiv \sum_i^{N}{\bf S}_i,
\end{equation}
where we exploit site independence of the global photon field, and
$N\equiv 2S$ plays the role of a large spin.  This motivates an
asymptotically exact semiclassical treatment based on the
Holstein--Primakoff transformation
\cite{Hillery:Semi,Persico:Coherence,Ressayre:HP}
\begin{equation}
\begin{aligned}
J^+ & = c^\dagger (2S-c^\dagger c)^{1/2},\\ 
J^- & = (2S-c^\dagger c)^{1/2}c,\\
J^z & = c^\dagger c-S,
\end{aligned}
\end{equation}
where $c$ is a canonical boson. The superradiance transition is
associated with condensation of this auxiliary boson. Most crucially,
this is related to condensation of {\em polaritons}, and {\em not} the
$a,b$ bosons themselves.  The Dicke model (\ref{DHam}) becomes
\begin{equation}
\begin{aligned}
H & =\tilde\omega_0 \left(c^\dagger c-S\right) 
+\tilde\omega\psi^\dagger\psi\\
& +\frac{\bar {\mathcal G}}{\sqrt{N}}\left(\psi^\dagger(2S-c^\dagger c)^{1/2}c 
+c^\dagger (2S-c^\dagger c)^{1/2}\psi\right),
\end{aligned}
\end{equation}
where we drop the `constant' $c_n$. Within a semiclassical $1/S$
expansion around the thermodynamic limit \cite{Hillery:Semi} we may
proceed by introducing coherent states for the photon and auxilliary
boson
\begin{equation}
|\gamma\rangle  \equiv e^{-\frac{\gamma^2}{2}+\gamma\psi^\dagger}|0\rangle,
\quad
|\zeta\rangle  \equiv e^{-\frac{\zeta^2}{2}+\zeta c^\dagger}|0\rangle,
\end{equation}
where $\gamma$ and $\zeta$ are classical $c$-numbers. The energy density reads
\begin{equation}
{\mathcal E}\equiv \langle H\rangle /N=\tilde\omega_0(\bar\zeta^2-1/2)
+\tilde\omega\bar\gamma^2
+2\bar {\mathcal G}\bar\gamma\bar\zeta(1-\bar\zeta^2)^{1/2},
\label{eden}
\end{equation}
where $\bar\gamma\equiv \gamma/\sqrt{N}$, $\bar\zeta\equiv
\zeta/\sqrt{N}$.  Minimizing over $\bar\gamma$ gives
\begin{equation}
\bar\gamma=
-\frac{\bar {\mathcal G}\bar\zeta(1-\bar\zeta^2)^{1/2}}{\tilde\omega}.
\label{gamzeta}
\end{equation}
Substituting back in (\ref{eden}) yields
\begin{equation}
{\mathcal E}=\frac{\bar {\mathcal G}^2}{\tilde\omega}\bar\zeta^4
+\left(\tilde\omega_0-
\frac{\bar {\mathcal G}^2}{\tilde\omega}\right)\bar\zeta^2
-\frac{\tilde\omega_0}{2},
\label{ezeta}
\end{equation}
and we acquire an expectation value $\langle c\rangle \neq 0$ when the
quadratic term becomes negative. This corresponds to the superradiance
quantum phase transition at $\bar {\mathcal
  G}_c=\sqrt{\tilde\omega\tilde\omega_0}$ in agreement with the
previous results. Minimizing (\ref{ezeta}) over $\bar\zeta^2$ and
substituting into (\ref{gamzeta}) one obtains
\begin{equation}
\bar\zeta^2=\frac{\bar {\mathcal G}^2-
\bar {\mathcal G}_c^2}{2\bar {\mathcal G}^2},\quad \bar\gamma^2
=\frac{\bar {\mathcal G}^4
-\bar {\mathcal G}_c^4}{4\tilde\omega^2\bar {\mathcal G}^2}; 
\quad \bar {\mathcal G}\ge \bar {\mathcal G}_c.
\label{holsteinonsets}
\end{equation}
The onset of the photon field agrees with equation (\ref{photonset}).
In addition, the expectation value $\langle J^+\rangle =
\zeta(2S-\zeta^2)^{1/2}$ tracks the condensation of the
Holstein--Primakoff boson, $c$, via its coherent state parameter
$\zeta$:
\begin{equation}
\frac{\langle J^+\rangle}{N}=\bar\zeta(1-\bar \zeta^2)^{1/2}
=-\frac{\tilde\omega\bar\gamma}{\bar {\mathcal G}}.
\label{jplusave}
\end{equation}
These results show that the superradiance transition is accompanied by
condensation of the Holstein--Primakoff boson, $\langle c\rangle\neq
0$. As we will discuss in section \ref{Sect:Coherence}, in the
two-band Bose--Hubbard model coupled to light, this corresponds to
condensation of the polaritonic {\em bilinear} $\langle b^\dagger
a\rangle$ above the Mott background, and {\em not} the $a$, $b$ bosons
themselves.  In addition to the onset of photons described by
equations (\ref{photonset}) and (\ref{holsteinonsets}), the
magnetization ${\mathcal M}\equiv \langle J^z\rangle/N$ reproduces the
previous results.

The correspondence between the variational approach outlined in
section \ref{Sect:Zerovar}, and the Dicke model analysis is clearly
encouraging. In section \ref{Sect:Hopping} we shall extend the
variational approach to include the important effects of itinerancy
and carrier superfluidity.
 
\section{Variational Approach for Hardcore Atoms}
\label{Sect:Hopping}
\subsection{Phase Diagram}

\begin{figure}
\begin{center}
\includegraphics[width=8cm]{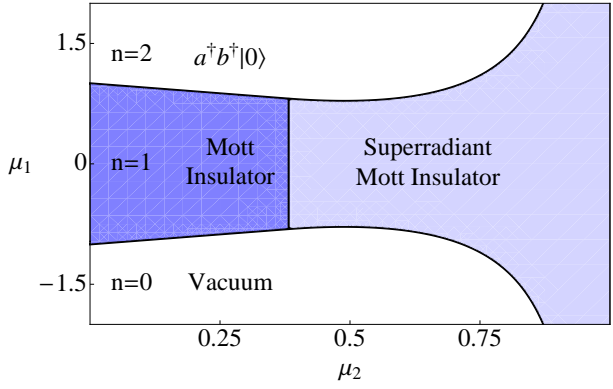}
\end{center}
\caption{(Color online). Variational zero hopping 
diagram for {\em hardcore} $a$
and $b$ atoms. The symmetry about $\mu_1=0$ reflects the combined
particle--hole and species interchange symmetry of the Hamiltonian
(\ref{reduced}) in the hardcore limit; see text.}
\label{Fig:varzhhc}
\end{figure}

Having confirmed a zero hopping Mott phase, with $n_a+n_b=1$, we
consider itinerancy and carrier superfluidity.  Within this lowest
lobe we may take {\em hardcore} $a$ {\em and} $b$ bosons.  This will
also be convenient for the numerical simulations in section
\ref{Sect:Numerics}.  Whilst this does not affect physics {\em within}
the lobe, the zero hopping upper boundary is modified by the
restriction on the $b$-atom population. In this case we only need
retain the states $|0,0\rangle$, $|1,1\rangle$ and the admixtures of
$|0,1\rangle$ and $|1,0\rangle$. The zero hopping diagram shown in
Fig.~\ref{Fig:zerohop} is replaced by Fig.~\ref{Fig:varzhhc}.  The
lower boundary remains unchanged because the same eigenstates are
involved, and the upper boundary becomes a mirror reflection of the
lower one.  This is quite natural since the $a$ and $b$ operators now
appear on an equal footing, modulo the effects of the band
splitting. More formally, this may be traced to the invariance of the
hardcore Hamiltonian (\ref{gham}) (up to a constant term) under the
particle--hole transformation $a\rightarrow a^\dagger$, $b\rightarrow
b^\dagger$, together with $\mu_1\rightarrow
\epsilon_a+\epsilon_b+U_{ab}-\mu_1$, and the interchange of the $a$
and $b$ operators and the hopping parameters $J_a$ and $J_b$; since
hardcore bosons obey onsite {\em anticommutation} relations,
$n\rightarrow 1-n$, under particle--hole transformation. For
$\epsilon_a=-1$, $\epsilon_b=1$ and $U_{ab}=0$ this involves
$\mu_1\rightarrow -\mu_1$. The particle--hole transformation also
accounts for the change in the vacuum, $|0,0\rangle\rightarrow
|1,1\rangle$.

\begin{figure}
\includegraphics[clip=true,width=7.5cm]{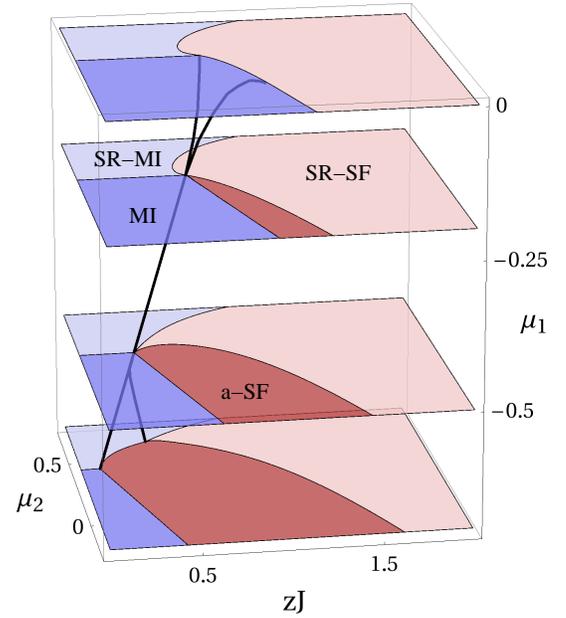}
\caption{(Color online). Variational phase diagram with $J_a=J_b=J$ and
$\epsilon_a=-1$, $\epsilon_b=1$, $\omega=\bar g=1$, $U_{ab}=0$. The
phases are (i) a Mott insulator (MI, dark blue), (ii) a superradiant Mott
state supporting a condensate of photo-excitations (SR-MI, light blue), (iii)
a superradiant superfluid (SR-SF, light red), and (iv) an a-type superfluid
(a-SF, dark red). As shown by the solid line, the tetracritical line
ultimately bifurcates into two bicritical points connected by a first
order transition. For $\mu_1>0$, the a-type superfluid is replaced by
a b-type owing to the particle--hole and species interchange symmetry
of the hardcore Hamiltonian (\ref{gham}) when $J_a=J_b$; see text.}
\label{Fig:pd}
\end{figure}

To incorporate itinerancy and superfluidity we augment the variational
analysis for two component bosons in an optical lattice
\cite{Altman:Twocomp} with a coherent state for light:
\begin{equation}
\begin{aligned}
|{\mathcal V}\rangle & = 
|\gamma\rangle\otimes\prod_i\left[\cos\theta_i(\cos\chi_i
  a_i^\dagger
+\sin\chi_i b_i^\dagger)\right.\\
& \hspace{1.6cm}
\left. +\sin\theta_i(\cos\eta_i+\sin\eta_i b_i^\dagger a_i^\dagger)\right]
|0\rangle,
\end{aligned}
\label{varstate}
\end{equation}
where $|\gamma\rangle$ is the coherent state introduced previously,
and $\theta,\chi,\eta,\gamma$ are to be determined. The corresponding
order parameters are given by $\langle a\rangle =\frac{1}{2}\sin
2\theta \cos(\chi-\eta)$, $\langle b\rangle =\frac{1}{2}\sin 2\theta
\sin(\chi+\eta)$ and $\langle \psi\rangle = \gamma$.  The first term
in brackets in equation (\ref{varstate}) describes the Mott state, and
the second superfluidity.  For $\theta=0$ this coincides with the
variational approach for localized excitons coupled to
light~\cite{Littlewood:Models}, and as we will discuss in section
\ref{Sect:Coherence}, reproduces the previous results for
$J_\alpha=0$. More generally, (\ref{varstate}) takes real hopping into
account, involving site vacancies and interspecies double
occupation. It provides a useful starting point to identify the
boundaries between the Mott and superfluid regions.  We consider
spatially uniform phases with energy density ${\mathcal
  E}\equiv\langle {\mathcal V}|H|{\mathcal V}\rangle/N$:
\begin{equation}
\begin{aligned}
{\mathcal E}& =(\tilde\epsilon_+ -\tilde\epsilon_-\cos 2\chi)
\cos^2\theta+(2\tilde\epsilon_++U_{ab})\sin^2\eta\sin^2\theta\\ &
-\frac{z}{4}\left[J_a\cos^2(\chi-\eta)+J_b\sin^2(\chi+\eta)\right]
\sin^2 2\theta\\ & +\tilde\omega\bar\gamma^2+\bar g\bar
\gamma\cos^2\theta\sin2\chi,
\end{aligned}
\label{varen}
\end{equation}
where $z$ is the coordination and $\tilde\epsilon_\pm\equiv
(\tilde\epsilon_b\pm\tilde\epsilon_a)/2$. Minimizing on $\bar\gamma$
gives $\bar\gamma=\langle \psi\rangle/\sqrt N= -\bar g\cos^2\theta\sin
2\chi/2\tilde\omega$. This may be eliminated from (\ref{varen}) to
yield
\begin{equation}
\begin{aligned}
{\mathcal E} = & (\tilde\epsilon_+ -\tilde\epsilon_-\cos 2\chi)
\cos^2\theta+(2\tilde\epsilon_++U_{ab})\sin^2\eta\sin^2\theta\\
& -\frac{z}{4}\left[J_a\cos^2(\chi-\eta)+J_b\sin^2(\chi+\eta)\right]
\sin^2 2\theta\\
&  -\frac{\bar g^2\cos^4\theta\sin^2 2\chi}{4\tilde\omega}.
\end{aligned}
\label{varen2}
\end{equation}
The symmetries of (\ref{varen2}) determine the domain of minimization.
It is invariant under $\theta\rightarrow \theta+\pi$ and
$\theta\rightarrow -\theta$, and may be minimized over $\theta\in
[0,\pi/2]$. Likewise, it is invariant under $\chi\rightarrow \chi+\pi$
and $\eta\rightarrow \eta+\pi$. Since the hopping contribution favors
$\chi$ and $\eta$ taking the same sign, we may also minimize
$\chi,\eta\in [0,\pi/2]$. To begin with we set $J_a=J_b=J$ and the
expression may be further reduced using
$J_a\cos^2(\chi-\eta)+J_b\sin^2(\chi+\eta)=J(1+\sin 2\chi\sin 2\eta)$.
Minimizing over the restricted domain yields the phase diagram in
Fig.~\ref{Fig:pd}.
For the chosen parameters, we have up to four distinct phases in the
interval $\mu_1<0$; (i) a Mott state with $\langle a\rangle=\langle
b\rangle =\langle\psi^\dagger\psi\rangle=0$, (ii) a superradiant Mott
state with $\langle a\rangle=\langle b\rangle=0$ and
$\langle\psi^\dagger\psi\rangle\neq 0$, (iii) a single component
superfluid with $\langle a\rangle\neq 0$ and $\langle
b\rangle=\langle\psi^\dagger\psi\rangle=0$, and (iv) a superradiant
superfluid $\langle a \rangle\neq 0$, $\langle b\rangle\neq 0$,
$\langle \psi^\dagger\psi\rangle\neq 0$. Indeed, the Hamiltonian
displays a ${\rm U}(1)\times {\rm U}(1)$ symmetry and these may be
broken independently. The phase diagram reflects this pattern of
symmetry breaking.  In particular, the superradiant Mott state
corresponds to an unbroken ${\rm U}(1)$ in the matter sector
(corresponding to a pinned density and phase fluctuations) but a
broken ${\rm U}(1)$ (or phase coherent condensate) for
photo-excitations. As we shall discuss below, the expectation value of
the {\em bilinear}, $\langle b^\dagger a\rangle \neq 0$, corresponds
to the onset of coherence in the Dicke model. This novel phase may
thus be regarded as a form of supersolid \cite{Andreev:SS} in which
photo-excitations condense on the background of a Mott
insulator. Although the lattice precludes {\em spontaneous}
translational symmetry breaking (at least with this periodicity) the
excitations may be thought of as mobile defects in an otherwise
ordered background.

As in the zero hopping case, we may extend Fig.~\ref{Fig:pd} into the
region $\mu_1>0$ by exploiting symmetries of (\ref{gham}). This is
reflected in the variational energy by using
$\sin^2\eta=(1-\cos2\eta)/2$ to combine the $\tilde\epsilon_+$
contributions:
\begin{equation}
\begin{aligned}
{\mathcal E} & = \frac{1}{2}(\epsilon_a+\epsilon_b)-\mu_1 -
\frac{1}{2}\left(\epsilon_b-\epsilon_a-\mu_2\right)\cos 2\chi
\cos^2\theta \\
& +\frac{1}{2}\left[U_{ab}-(\epsilon_a+\epsilon_b+U_{ab}-2\mu_1)
\cos 2\eta\right]\sin^2\theta\\
& -\frac{zJ}{4}\left(1+\sin 2\chi\sin 2\eta\right)\sin^2 2\theta
 -\frac{\bar g^2\cos^4\theta\sin^2 2\chi}{4(\omega-\mu_2)}.
\end{aligned}
\end{equation}
Since $J_a=J_b=J$ this is invariant (up to a constant) under 
$\mu_1\rightarrow \epsilon_a+\epsilon_b+U_{ab}-\mu_1$ and
$\eta\rightarrow \pi/2-\eta$, which interchanges the 
superfluid order parameters $\langle a\rangle$
and $\langle b\rangle$. For the parameters used in Fig.~\ref{Fig:pd},
the $a$-type superfluid observed for $\mu_1<0$ is replaced with a
$b$-type for $\mu_1>0$.

As may be seen from Fig.~\ref{Fig:pd}, the locus of the tetracritical
line may be determined from the intersection of the superradiance
transition with the Mott insulator to a-type superfluid phase
boundary.  This yields $\tilde\epsilon_a+zJ=0$ as discussed in
Appendix \ref{App:varpb}.  This may also be seen from a Landau type
expansion of ${\mathcal E}$. In view of the non-linear relationship,
it is convenient to expand in the angles $\theta,\chi,\eta$ as opposed
to the order parameters $\langle a\rangle$, $\langle b\rangle$,
$\langle\psi\rangle$:
\begin{equation} {\mathcal
    E}=\tilde\epsilon_a-(\tilde\epsilon_a+zJ)\theta^2
  -\left(\frac{\bar
      g^2-\tilde\omega\tilde\omega_0}{\tilde\omega}\right)\chi^2+\dots
\end{equation}
The quadratic ``mass'' terms vanish when $\tilde\epsilon_a+zJ=0$ and
the superradiance condition $\bar g=\sqrt{\tilde\omega\tilde\omega_0}$
is met. In some simple cases ${\mathcal E}$ may be expanded directly
in the order parameters. For example, throughout the entire Mott lobe
where $\langle a\rangle=\langle b\rangle=0$ we may set
$\theta=0$. Returning to equation (\ref{varen}) and minimizing over
$\chi$ yields $\tan 2\chi=-2\bar g\bar\gamma/\tilde\omega_0$. This
gives ${\mathcal E}=\tilde\epsilon_++\tilde\omega\bar \gamma^2
-\sqrt{\tilde\omega_0^2/4+\bar g^2\bar \gamma^2}$, where
$\bar\gamma=\langle \psi\rangle/\sqrt{N}$. This agrees with our
variational zero hopping result (\ref{enminus}) in the lowest lobe
with $n=1$.

\begin{figure}
\begin{center}
\includegraphics[width=8cm]{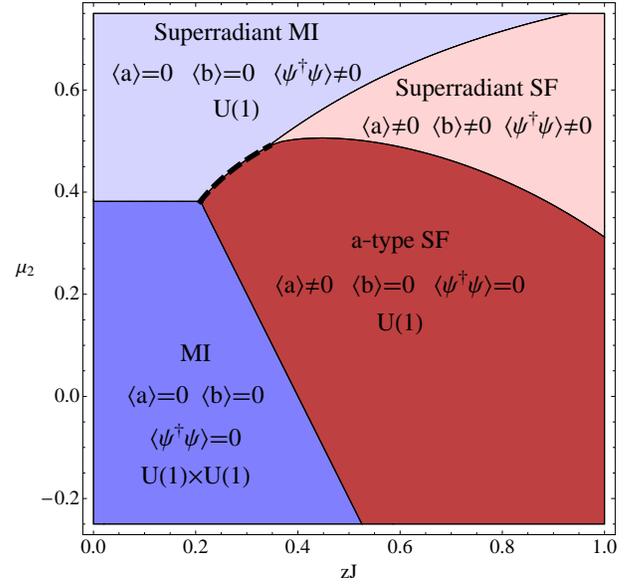}
\caption{(Color online). Cross section of the phase diagram
  (\ref{Fig:pd}) for $\mu_1=-0.6$.  The first order transition from
  the superradiant Mott state to the $a$-type superfluid is indicated
  by a dashed line. The locus of this transition is determined
  analytically from (\ref{SRMIASF}).  The remaining transitions are
  continuous.\label{Fig:Firstorder}}
\end{center}
\end{figure}

\begin{figure}
\begin{center}
\includegraphics[width=8cm,clip=true]{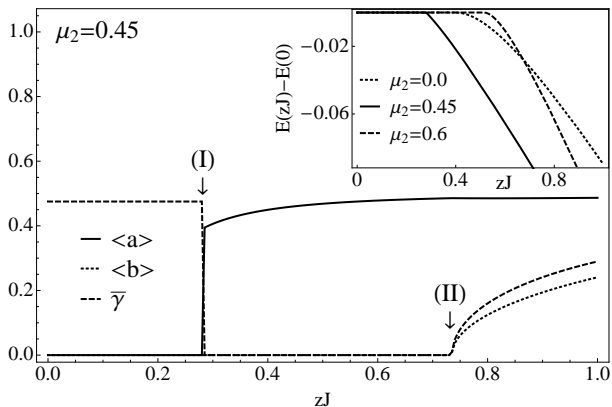}
\caption{Evolution of the mean field order parameters $\langle
a\rangle$, $\langle b\rangle$,
$\bar\gamma=\langle\psi\rangle/\sqrt{N}$ across the transitions
depicted in Fig.~\ref{Fig:Firstorder} for $\mu_2=0.45$. (I) First
order transition from the superradiant Mott state to the $a$-type
superfluid, (II) continuous transition from the $a$-type superfluid to
the superradiant superfluid. 
Inset: Evolution of the variational energy (relative to the zero
hopping Mott phase) across the transitions in
Fig.~\ref{Fig:Firstorder} for fixed values of $\mu_2$. The solid line
at $\mu_2=0.45$ has a discontinuity in the first derivative and
indicates a first order transition in passing from the superradiant
Mott insulator to the $a$-type superfluid.  The remaining transitions
are continuous.
\label{Fig:ordersenergy}}
\end{center}
\end{figure}

\begin{figure}
  \includegraphics[clip=true,width=7.5cm]{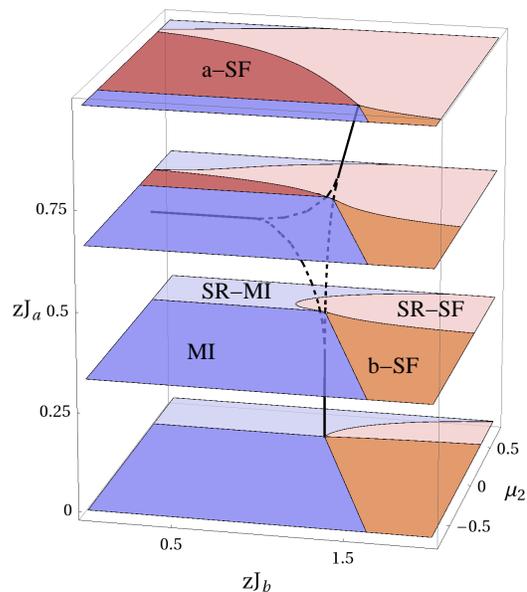}
  \caption{(Color online). Variational phase diagram showing a slice
    through Fig.~\ref{Fig:varzhhc} with $\mu_1=-0.25$ and extended in
    to the $(J_a,J_b)$ hopping plane. We use the same key 
 as in Fig.~\ref{Fig:pd} and denote the $b$-type superfluid
    (b-SF) in orange. Single component superfluids are observed for
    large hopping asymmetry.}
\label{Fig:pdjajb}
\end{figure}

In Fig.~\ref{Fig:Firstorder} we present a cross section of the phase
diagram (\ref{Fig:pd}) for $\mu_1=-0.6$.
We indicate continuous transitions by single lines and first order
transitions by dotted lines. In general, the transition from the
superradiant Mott state to the non-superradiant $a$-type superfluid
involves a discontinuous jump in the photon density.  This results in
a first order transition as indicated by the discontinuity in the
first derivative of the energy and the order parameters as shown in
Fig.~\ref{Fig:ordersenergy}.
This first order line connects two bicritical points as indicated in
Fig.~\ref{Fig:Firstorder}. The length of this first order segment
changes with $\mu_1$, and emanates from the tetracritical line as
shown in Fig.~\ref{Fig:pd}.  These overall features are also exhibited
for more general hopping parameters as shown in Fig.~\ref{Fig:pdjajb},
where $a$-type and $b$-type superfluids emerge for large hopping
asymmetry.

\subsection{Coherence in Superradiant Mott Phase}
\label{Sect:Coherence}
An interesting aspect of the superradiant Mott phase is that polariton
condensation coexists with the Mott character.  Within both Mott
phases $\theta=0$ and our variational wavefunction (\ref{varstate})
becomes
\begin{equation}
|{\mathcal V}\rangle=|\gamma\rangle\otimes \prod_i
\left(\cos\chi_ia_i^\dagger +\sin\chi_i b_i^\dagger\right)|0\rangle.
\label{thetzerovar}
\end{equation}
To describe photo-excitations above the filled Mott state, we
introduce a change of vacuum 
$|\Omega\rangle \equiv\prod_i a_i^\dagger |0\rangle$
so that
\begin{equation}
|{\mathcal V}\rangle=|\gamma\rangle\otimes \prod_i
\cos\chi_i\prod_i\left(1 +\tan\chi_i b_i^\dagger
a_i^{\phantom\dagger}\right)|\Omega\rangle.
\end{equation}
Since we are dealing with hardcore bosons this may be exponentiated,
and for homogenous parameters
\begin{equation}
|{\mathcal V}\rangle=|\gamma\rangle\otimes e^{N\ln
  \cos\chi}e^{\tan\chi \sum_i b_i^\dagger
  a_i^{\phantom\dagger}}|\Omega\rangle.
\end{equation}
This is already reminiscent of a coherent state for the {\em
bilinears}, although one needs to be careful since we are dealing with
hardcore bosons.  Instead, we may examine condensation properties
directly by computing expectation values using (\ref{thetzerovar}).
Using the Schwinger boson representation (\ref{hcorespins}) for the
lowest lobe
\begin{equation}
{\mathcal J}^+\equiv \sum_i b_i^\dagger a_i^{\phantom\dagger},\quad 
{\mathcal J}^-\equiv \sum_i a_i^\dagger b_i^{\phantom\dagger},\quad
{\mathcal J}^z\equiv \frac{1}{2}\sum_i (n^b_i-n^a_i),
\end{equation}
we may calculate the bosonic bilinear $\langle {\mathcal J}^+\rangle$:
\begin{equation}
\langle {\mathcal V}|{\mathcal J}^+|{\mathcal V}\rangle = \sum_i
\langle {\mathcal V} |b_i^\dagger a_i^{\phantom\dagger}|{\mathcal
  V}\rangle =N\sin \chi\cos \chi=\frac{N}{2}\sin 2\chi.
\end{equation}
At $\theta=0$ our variational analysis yields
$\bar\gamma=-\bar g\sin 2\chi/2\tilde\omega$:
\begin{equation}
\frac{\langle {\mathcal V}|{\mathcal J}^+|{\mathcal V} \rangle}{N}
\equiv \frac{\sum_i \langle b_i^\dagger
a_i^{\phantom\dagger}\rangle}{N}= -\frac{\tilde\omega\bar\gamma}{\bar
g}.
\end{equation}
This agrees with the result (\ref{jplusave}) obtained from the
Holstein--Primakoff approach to the Dicke model \cite{Hillery:Semi}.
We see that in the two-band Bose--Hubbard problem we have condensation
involving {\em particle-hole pairs} above the Mott background with
$\langle b^\dagger a\rangle \neq 0$.  This corresponds to condensation
of the Holstein--Primakoff boson $\langle c\rangle \neq 0$ in the dual
formulation as evidenced by equation (\ref{jplusave}).  Crucially,
condensation of $\langle b^\dagger a\rangle $ is {\em not} accompanied
by condensation of the individual bosonic carriers $a$ and $b$. This
follows immediately from (\ref{thetzerovar}) where $\langle
a\rangle=\langle b\rangle=0$. Since matter--light coherence in the
Dicke model translates into polariton and not carrier condensation,
Mott behavior and polariton condensation may coexist.

\section{Numerical Simulations}
\label{Sect:Numerics}
We now analyze the Hamiltonian (\ref{qlmodel}) by exact
diagonalization, imposing a maximum number of photons,
$\las\psi^\dag\psi\ras\leq M_\psi$, in addition to the hardcore $a,b$
constraints. We consider a one-dimensional system with $N$ lattice
sites, with the basis of tensor product states
$\ket{\phi}_a^{(\nu_a)}\otimes \ket{\phi}_b^{(\nu_b)}
\otimes\ket{\phi}_\psi^{(\nu_\psi)}$, where $\nu_{a,b}=1,\dots,2^N$
and $\nu_\psi=0,1,\dots,M_\psi$. Here
$\ket{\phi}^{(\nu_\alpha)}_\alpha=
\ket{n_{\alpha,1}^{(\nu_\alpha)},\dots,n_{\alpha,N}^{(\nu_\alpha)}}$
where $n^{(\nu_\alpha)}_{\alpha,i}\in\{0,1\}$, and the photon states
are given by $\ket{\phi}^{(\nu_\psi)}_\psi =
\ket{n^{(\nu_\psi)}_{\psi}}$ where
$n^{(\nu_\psi)}_{\psi}\in\{0,1,\dots,M_\psi\}$.  The total Hilbert
space has dimension $D=2^N 2^N(M_\psi+1)$.  In the largest case
considered, with $N=8$ sites and $M_\psi=64$ photons, this corresponds
to a matrix dimension $D\approx 4E6$.  Increasing $M_\psi$ further has
only minor influence.

Applying periodic boundary conditions in real space, we diagonalize
the sparse matrix representation of $H$ to obtain the groundstate
$\ket{\Phi_0}$ and its energy $E_0$.  We compute the atom and photon
densities, and the density fluctuations $\sigma_{\alpha} = \sqrt{\las
  \hat{n}_{\alpha}^2\ras - \las \hat{n}_{\alpha}\ras^2}$. To obtain
the superfluid fraction, $f_s^\alpha$, of the $\alpha$ atoms, we
impose a phase twist $\Theta\ll\pi$ by means of a Peierls factor
$\alpha^\dagger_i \alpha^{\nag}_j\mapsto \alpha^\dagger_i
\alpha^{\nag}_j e^{-\rmi\Theta/N}$, and calculate the change in the
ground state energy \cite{Roth:Twocolor}:
\begin{equation}
  f^\alpha_s = \frac{N}{J_\alpha \las n_\alpha \ras}
  \frac{E^{(\Theta)}_0-E_0}{\Theta^2}.
\end{equation}
For the single-component Bose--Hubbard model, this quantity is zero
deep in the Mott insulator, and approaches unity far in the
superfluid.  Note that in our case, the hardcore constraint always
provides an effective interaction even for large hopping, so that
$f_s<1$. A total superfluid density can be obtained by imposing the
phases on both species a,b and calculating
\begin{equation}
f_s = \frac{N}{\sum_\alpha J_\alpha \las n_\alpha \ras}
\frac{E^{(\Theta)}_0-E_0}{\Theta^2} \,.
\label{fs}
\end{equation}
We supplement these superfluid diagnostics with the zero
momentum occupations
$ n_\alpha(k=0) = N^{-1}\sum_{pq} \las \alpha^\dag_p
 \alpha^\nag_q \ras$.

\subsection{Small Hopping and Dicke Superradiance}

\begin{figure}
\begin{center}
\includegraphics[width=7.7cm,clip=true]{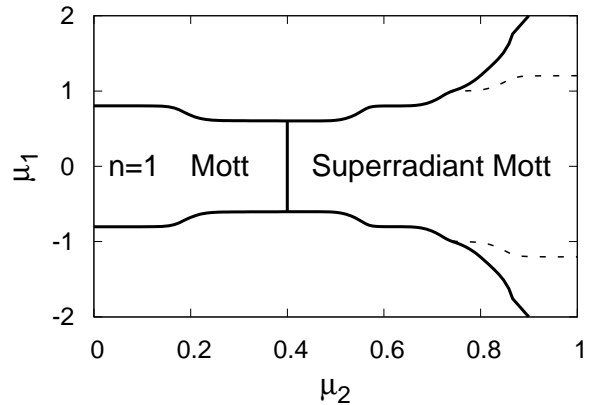}
\caption{Exact diagonalization results for the phase diagram at small
  hopping, $zJ=0.1$, for {\em hardcore} $a$,$b$ atoms and
  $\epsilon_a=-1$, $\epsilon_b=1$, $\omega=\bar g=1$, $U_{ab}=0$. The
  dashed (solid) lines corresponds to $N=8$ and $M_\psi=16$
  ($M_\psi=64$) photons, and have been obtained from spline fits to
  the contours at $n=0.99$ and $n=1.01$.  The vertical line indicates
  the location of the superradiance transition as determined from
  $\las \psi^\dagger\psi\ras/N\ge 0.01$.  The results are in good
  qualitative agreement with the variational zero hopping results in
  Fig.~\ref{Fig:varzhhc}, showing the continuity of this behavior into
  the Mott phase.}
\label{Fig:ZHHC}
\end{center}
\end{figure}

\begin{figure}
\begin{center}
  \includegraphics[width=6.9cm,clip=true]{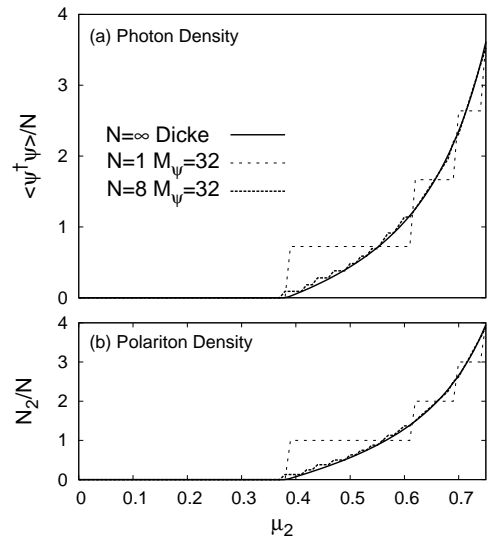}
\caption{Exact diagonalization results with $N=1,8$ sites, $J_a=J_b=J$
and $zJ=0.1$.  We set $\epsilon_a=-1$, $\epsilon_b=1$, $\omega=\bar
g=1$, $\mu_1=U_{ab}=0$.  The profiles show the onset of superradiance
within the Mott phase, and evolve with increasing system size towards
those for the thermodynamic limit of the Dicke model.  Evolution
of (a) the photon density $\langle \psi^\dagger \psi\rangle/N$,  (b)
the polariton density, $N_2/N$.  For $N=1$ the Dicke
model reduces to the Jaynes--Cummings model and exhibits quantized
integer steps in the polariton density
\cite{Greentree:QPTlight,Koch:JCL}. More generally these are quantized
in units of $1/N$. The value of $M_\psi$ is of minor importance in the
range of $\mu_2$ considered as it sets the upper limit on the number
of excitations available.}
\label{Fig:Finiteonset}
\end{center}
\end{figure}

To verify the existence of a superradiance transition {\em within} the
Mott phase we begin our numerical treatment in the limit of small
non-zero hopping.  Figure~\ref{Fig:ZHHC} shows the Mott lobe with
density $n=1$ at $zJ=0.1$, for two different values of $M_\psi$. For
sufficiently large $M_\psi$ the results are in good agreement with the
variational {\em zero-hopping} phase diagram shown in
Fig.~\ref{Fig:varzhhc}, and the overall features depend only weakly on
the number of sites. In order to examine the onset of superradiance,
we track the evolution of the polariton and photon densities in
Fig.~\ref{Fig:Finiteonset}. At zero hopping, equation (\ref{BGC})
yields the critical coupling
$\overline{g}^2_\text{c}=\tilde{\om}\tilde{\om}_0
=(\om-\mu_2)(\en_b-\en_a-\mu_2)$.  For the chosen parameters at fixed
$\bar g=1$, the transition occurs at a critical chemical potential,
$\mu_2^c=(3-\sqrt{5})/2\approx 0.382$. This onset is well reproduced
at small finite hopping, as shown in Fig.~\ref{Fig:Finiteonset}. For
$N=1$ (where the Dicke model reduces to the Jaynes--Cummings model) we
see quantized steps with fixed integer polariton density, $N_2/N$.
These form the basis of the Mott lobes observed in
Jaynes--Cummings--Hubbard models
\cite{Greentree:QPTlight,Koch:JCL}. Even in this extreme limit, the
densities closely track the thermodynamic Dicke model results.  As $N$
increases, the step sizes are reduced by a factor of $1/N$ and we
approach the variational thermodynamic results.  This behavior in our
itinerant boson model closely mirrors direct finite $N$ simulations of
the Dicke model; see Fig.~\ref{Fig:FiniteDicke}. A notable difference
between the Dicke and Jaynes--Cummings--Hubbard models
\cite{Greentree:QPTlight,Koch:JCL} (both with $N$ sites) is reflected
in their zero hopping eigenstates. In the latter, eigenstates are
tensor products of superpositions of two states which differ in photon
number by one, whereas in the Dicke model the global photon mode leads
to a coherent photon state for large $N$.  In the Dicke model the
higher Mott lobes are eliminated in favor of a continuous photon
onset, whereas in the Jaynes--Cummings--Hubbard models
\cite{Greentree:QPTlight,Koch:JCL} they remain in tact.

\subsection{General Phase Diagram}

\begin{figure}
\includegraphics[width=7.2cm,clip=true]{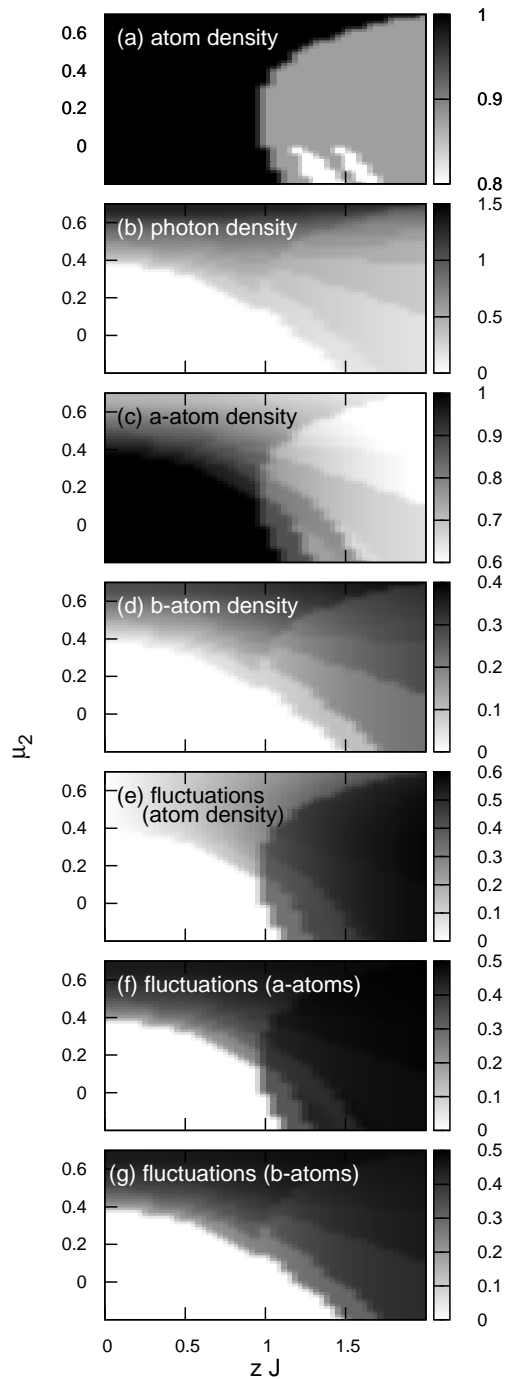}
\caption{Exact diagonalization results for $N=8$ sites and $M_\psi=16$
photons. We set $\mu_1=0$, $J_a=J_b=J$ and $\epsilon_a=-1$,
$\epsilon_b=1$, $\omega=\bar g=U_{ab}=1$.  The panels show density
plots of (a) total atom density, (b) photon density, (c) a-atom
density, (d) b-atom density, (e) fluctuations of the total atom
density, (f) fluctuations of the a-atom density, (g) fluctuations of
the b-atom density. In particular (a) reveals the Mott--superfluid
transition and (b) the superradiance transition.}
\label{Fig:Panel}
\end{figure}

\begin{figure}
\includegraphics[width=7.5cm]{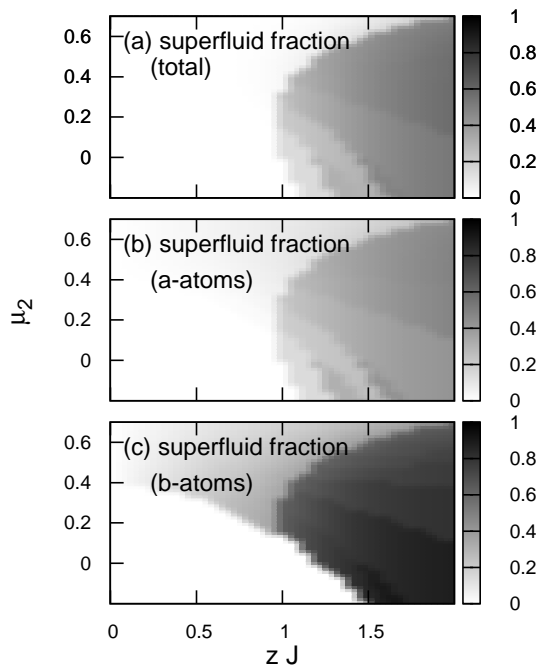}
\caption{Exact diagonalization results for $N=8$ sites and $M_\psi=16$
  photons. We set $\mu_1=0$, $J_a=J_b=J$ and $\epsilon_a=-1$,
  $\epsilon_b=1$, $\omega=\bar g=U_{ab}=1$.  The panels show density
  plots of (a) total atomic superfluid fraction, (b) a-atom superfluid
  fraction, (c) b-atom superfluid fraction. The Mott--superfluid
  transition is evident from the onset in panel (a), and the single
  component $a$-type superfluid in panels (b) and (c).}
\label{Fig:SFPanel}
\end{figure}

Turning to the construction of the overall phase diagram, we first
consider a fixed value $\mu_1=0$ and present results in the
$(zJ,\mu_2)$ plane; see Figures \ref{Fig:Panel} and
\ref{Fig:SFPanel}. So far we have set $U_{ab}=0$ in order to expose
the essential details.  Following our original work
\cite{Bhaseen:Polaritons}, we now take $U_{ab}=1$ to help illustrate
the generality of the overall results.  The interspecies interaction
also helps stabilize the Mott region up to larger values of the
hopping.
 
In Fig.~\ref{Fig:Panel} we show density plots of atom and photon
densities, and density fluctuations. In Fig.~\ref{Fig:SFPanel} we plot
superfluid fractions.  From these data, we can identify the four
different phases present in Fig.~\ref{Fig:pd}. The Mott--superfluid
transition is visible from the deviation of the total atom density
from unity in Fig.~\ref{Fig:Panel}\,(a). The superradiance transition
corresponds to the onset of photon density in
Fig.~\ref{Fig:Panel}\,(b). Panels (c) and (d) demonstrate the
existence of the single-component superfluid in the region where
$n\neq 1$ and $n_a\neq0$, while $n_b=0$.  This is also visible in the
individual superfluid fractions shown in Fig.~\ref{Fig:SFPanel}. We
also see the onset of a non-trivial population imbalance, $n_b-n_a$,
or existence of $b$-bosons, accompanying the superradiance
transition. Specifically, $n_a=1,n_b=0$ in the left of the white
region of panel (b), but $n_a<1,n_b>0$ above; see panels (c), (d) and
(g).  The nature of the two Mott phases (normal and superradiant) is
illustrated by panels (e)--(f).  We observe an onset of local
fluctuations $\sigma_\alpha$ with increasing hopping. Since the Mott
state is characterized by $n=n_a+n_b=1$, the individual atom density
fluctuations $\sigma_\alpha$ in the superradiant Mott phase are much
larger than $\sigma$. In the normal Mott phase, $n_b=0$ so that the
Mott insulator consists purely of a-atoms and hence
$\sigma_a=\sigma_b=0$. Finally, the Mott--superfluid transition is
also apparent in our results for the superfluid fractions as shown in
Fig.~\ref{Fig:SFPanel}.

\begin{figure}
\begin{center}
\includegraphics[width=7cm]{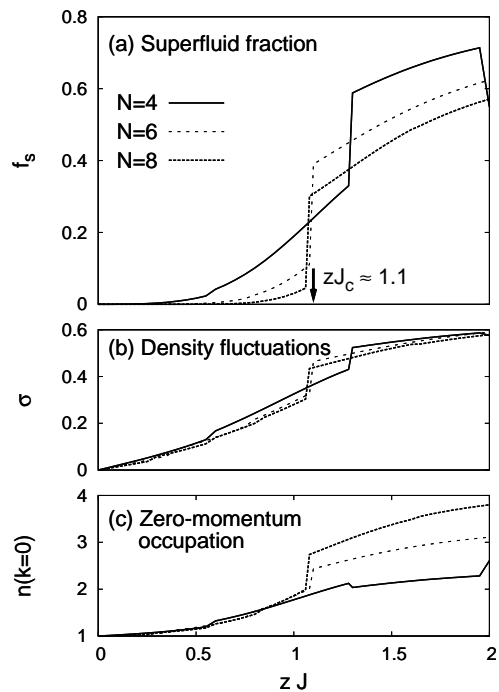}
\caption{Exact diagonalization results with $J_a=J_b=J$ and
$\epsilon_a=-1$, $\epsilon_b=1$, $\omega=\bar g=U_{ab}=1$.  We set
$\mu_1=0$, $\mu_2=0.4$, $M_\psi=2N$, and show the system size
dependence of (a) superfluid fraction, (b) density fluctuations, (c)
zero momentum occupation. Within the Mott phase for $J<J_c$, the
superfluid fraction decreases with increasing system size. This
indicates a Mott--superfluid transition at approximately $z J_c\approx
1.1$. This is compatible with the fluctuation onset criterion,
$\sigma=0.3$, that we use to determine the overall phase diagram in
Fig.~\ref{Fig:EDstack}, and the divergence of $n(k=0)$ with system
size in panel (c).
\label{Fig:Mottfluc}}
\end{center}
\end{figure}

Exact diagonalization of small clusters yields an approximation to the
critical value $J_c$ for the Mott-superfluid transition by monitoring
the onset of $f_s$, as in Fig.~\ref{Fig:SFPanel}\,(a), or the local
fluctuations $\sigma$ of the total atom density, as in
Fig.~\ref{Fig:Panel}\,(e).  Both $\sigma$ and $f_s$ are also nonzero
{\em below} $J_c$ for finite $N$ \cite{Roth:Twocolor}. While $f_s$ is
expected to scale to zero for $J<J_c$ as $N\to\infty$, the local
density fluctuations remain nonzero in the Mott phase due to virtual
hopping processes. This is illustrated in Fig.~\ref{Fig:Mottfluc}.
The system size $N$ has a noticeable impact on the onset of
correlations related to superfluidity.  Since a finite size scaling
analysis requires much larger system sizes, we provide approximate
phase boundaries obtained from the contour lines
$\sigma(zJ,\mu_2)=0.3$ (alternatively $f_s(zJ,\mu_2)=\mathrm{const.}$)
to indicate the onset of superfluidity, and
$\langle\psi^\dagger\psi\rangle/N=0.01$ to indicate superradiance. The
numerical constants for the contours have been chosen in order to
obtain phase boundaries that match those suggested by the data in
Figs.~\ref{Fig:Panel} and \ref{Fig:SFPanel}; see
Fig.~\ref{Fig:Mottfluc}.  Figure \ref{Fig:Mottfluc}~(c) shows the
zero-momentum occupation $n(k=0)=n_a(k=0)+n_b(k=0)$ for the atoms.
The latter is expected to diverge as a function of system size in the
SF phase, and such behavior can indeed be seen above $zJ\approx1.1$.

\begin{figure}
\begin{center}
\includegraphics[width=7.5cm]{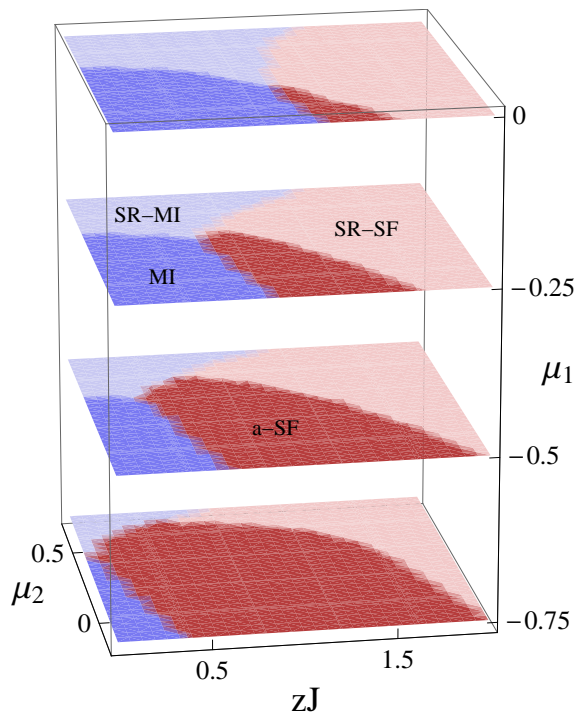}
\end{center}
\caption{(Color online). 
Overall phase diagram obtained by exact diagonalization with
$N=8$ sites and $M_\psi=16$ photons. We set $J_a=J_b=J$,
$\epsilon_a=-1$, $\epsilon_b=1$, $\omega=\bar g=1$, and $U_{ab}=1$,
and use the same  key as in Fig.~\ref{Fig:pd}.  We extract the
Mott--superfluid boundaries from the onset of density fluctuations,
$\sigma\ge 0.3$, and the superradiance transition from the onset of
photons, $\langle\psi^\dagger\psi\rangle/N\ge 0.01$; see text.  The
simulations show the distinct phases and the bifurcation of the
tetracritical line suggested by mean field theory. The bifurcation is
indicated by the regions where the two boundaries overlap along a line
rather than a single tetracritical point.}
\label{Fig:EDstack}
\end{figure}

Repeating the above procedure for different $\mu_1$ values we build up
a picture of the overall phase diagram as shown in
Fig.~\ref{Fig:EDstack}.  This may be compared to the variational
analysis shown in Fig.~\ref{Fig:pd}. We find the same phases as in the
analytical approach and the evolution of the phase boundaries is in
good agreement. For the choice of $\mu_1=0$ these phases meet in a
tetracritical point. As found analytically this extends into a
tetracritical line which ultimately bifurcates into two bicritical
points. The agreement between the numerical simulations and mean field
theory is remarkable given the enhanced role of fluctuations in low
dimensions. This mirrors the success of mean field theory in other
bosonic systems and may be assisted by the long range cavity photons.

\section{Relation To Other Problems}
\label{Sect:Other}

A feature not addressed by the present mean field theory, but captured
in Fig.~\ref{Fig:EDstack}, is the dispersion of the superradiance
transition with $J$; in the Mott phase, $\theta=0$, and $J$ drops out
of the variational energy (\ref{varen}).  One way to understand this
is to recast the matter contribution as
\begin{equation}
|{\mathcal V}_{\rm M}\rangle = \prod_i(\cos\chi_i +\sin\chi_i
b_i^\dagger a_i)|\Omega\rangle,
\end{equation}
where $|\Omega\rangle\equiv \prod_i a_i^\dagger |0\rangle$ is the filled 
Mott state in the absence of excitations. In the spatially uniform case
\begin{equation}
|{\mathcal V}_{\rm M}\rangle={\mathcal N}e^{\lambda\sum_i b_i^\dagger
 a_i}|\Omega\rangle,
\end{equation}
where $\lambda\equiv \tan\chi$ and ${\mathcal N}\equiv
(1+\lambda^2)^{-N/2}$.  This only accommodates {\em local}
particle-hole pairs above the filled Mott background, and may thus be
regarded as the BEC limit of the BEC--BCS crossover problem.  By
analogy with the fermionic BCS approach to exciton insulators
\cite{Keldysh:Kopaev,Keldysh:Kozlov}, and the crossover problem in
$^{40} {\rm K}$ \cite{Regal:BCSBEC}, one expects that excitons may
lower their energy by spreading out in real space and pairing in
momentum space. In this regard, the ``BCS'' pairing phenomenon in Bose
gases has a long history, with the Valatin--Butler wavefunction
\cite{Valatin:Collective} playing the role of the BCS state.  This has
been developed in a series of early works motivated by liquid $^ 4{\rm
  He}$ \cite{Cummings:TSF,Cummings:TSFErrata,Coniglio:Condensation,
  Coniglio:Coexistence,Evans:Bosepairing,Nozieres:Paircond} and
biexciton formation
\cite{Mavroyannis:Resonant,Mavroyannis:Coherent,Moskalenko:Book}. This
later emerged in studies of High-$T_c$ superconductivity
\cite{Rice:Superbose} and two-component Bose gases
\cite{Kagan:Pairing,Efremov:Two}. The interesting possibility of bound
states of three or four particles has also been explored
\cite{Brodsky:BS,Brodsky:Exact}.  Central to these studies is the
requirement to stabilize attractive Bose gases against collapse, with
the aid of short-range repulsion or internal structure. When this
condition is met, and at sufficiently low densities where exciton
overlap is negligible, it has been argued that such pairing states may
exist \cite{Nozieres:Paircond}.  In the present context, stabilization
of the paired state and the absence of carrier condensation is brought
about by the interplay of Mott physics and photoexcitation.  It would
be interesting to explore this problem in more detail, and we leave
these refinements to future studies.

As noted in our previous work, the connection to the BEC--BCS
crossover for bosons is reinforced by the Feshbach resonance problem
studied in the continuum
\cite{Rad:Atmol,Romans:QPT,Radzi:Resonant,Koetsier:Bosecross,Zhou:PS}
and on the lattice
\cite{Bhaseen:Feshising,Rousseau:Fesh,Rousseau:Mixtures}. Performing a
particle--hole transformation, the matter--light coupling reads
$\psi^\dagger a_i b_i$.  Aside from the global nature of the photon,
this converts $a$ and $b$ into a ``molecule'' $\psi$. At the outset
there are eight possible phases corresponding to separate condensation
of $\langle a\rangle$, $\langle b\rangle$, $\langle\psi\rangle$. At
the mean field level, only five of these may survive; condensation of
two variables provides an effective field (as dictated by the
coupling) which induces condensation of the other.  In Figures
\ref{Fig:pd} and \ref{Fig:EDstack} the band asymmetry,
$\epsilon_a<\epsilon_b$, in conjunction with the chosen parameters,
reduces this to four or less. Nonetheless, the additional b-type
superfluid is stabilized for larger values of $\mu_1$ owing to the
particle--hole and species interchange symmetry involving
$\mu_1\rightarrow \epsilon_a+\epsilon_b+U_{ab}-\mu_1$.  In contrast to
the single species mean field theory
\cite{Rad:Atmol,Romans:QPT,Radzi:Resonant}, this two species case
supports an atomic superfluid, since condensation of one carrier no
longer induces an effective field. Moreover, condensation may leave a
${\rm U}(1)$ symmetry intact, which allows the coexistence of Mott and
phase coherent behavior.

In deriving (\ref{varen}) and the phase diagram, we are primarily
concerned with the matter-light coupling. As such we incorporate
$U_{ab}$ as in Ref.~\onlinecite{Altman:Twocomp}. Within this
variational approach, this interaction has no impact on the Mott
states with $\theta=0$, as evident from the energy density
(\ref{varen}). Nonetheless, the presence of the matter-light coupling
stabilizes the non-trivial phases in Fig.~\ref{Fig:pd} and provides
good agreement with the numerical simulations. However, as noted by
S\"oyler {\em et al}~\cite{Soyler:Twocomp} and previous works
\cite{Kagan:Pairing,Efremov:Two,Brodsky:BS,Brodsky:Exact}, analogous
pairing phases may be supported in the two-component Bose--Hubbard
model, {\em without} matter-light coupling, through a more
sophisticated treatment of $U_{ab}$ itself. Indeed, onsite repulsive
interactions, $U_{ab}n_an_b$, favor a particle of one species and a
hole of the other on the same site. Treating this pairing in a BCS
approach, one may replace $n^a_i n^b_i$ by $|\Delta_i|^2+(\Delta_i
b_i^\dagger a_i +{\rm h.c.})$, where $\Delta_i\equiv \langle
a_i^\dagger b_i\rangle$, is to be determined self-consistently. This
field acts as a {\em local} ``photon'', and a similar mean field
phenomenology may ensue. Such pairing also occurs in fermionic models
\cite{Kantian:ALE}.  Although our discussion has focused on a single
{\em global} photon, the symmetry analysis is more general. This is
supported by studies of the two-band Bose--Hubbard model for equal
fillings and commensurate densities \cite{Kuklov:Comm}.

\begin{figure}
\begin{center}
\includegraphics[width=6cm,clip=true]{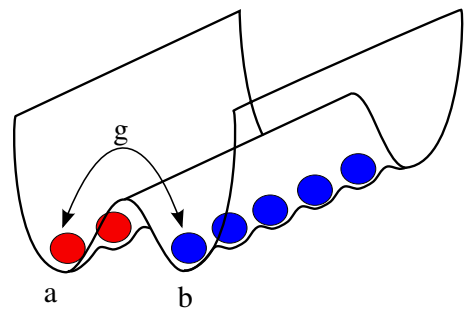}
\end{center}
\caption{(Color online.)  The classical light limit of equation
  (\ref{qlmodel}) (where $\psi$ is a replaced by a c-number), may be
  simulated in optical superlattices \cite{Jaksch:CBAOL} where the
  coupling $g$ represents tunnelling between different wells, $a$ and
  $b$.\label{Fig:Super}}
\end{figure}

In closing we note that the classical light limit of equation
(\ref{qlmodel}) (where $\psi$ is replaced by a c-number) may be
simulated in optical superlattices \cite{Jaksch:CBAOL} where
$g_ia_ib_i^\dagger$ represents tunnelling between different wells; see
Fig.~\ref{Fig:Super}.  In the case of hardcore $a$-atoms and softcore
$b$-atoms this provides an analogue of the Jaynes--Cummings--Hubbard
models considered in Refs.~\cite{Greentree:QPTlight,Koch:JCL}. This
geometry may also be useful in realizing other ``matter--bath''
systems.

\section{Conclusions}
\label{Sect:Conc}
We have investigated the two-band Bose--Hubbard model coupled to a
single mode of a cavity light field. We combine analytical and
numerical techniques and find good agreement between the approaches.
The model displays a novel phase in which ``polaritons'' condense on
the platform of a bosonic Mott insulator.  We extend our previous work
\cite{Bhaseen:Polaritons} in several directions including an
investigation of the overall phase diagram, and the nature of
polariton condensation.  In particular, we use the framework of the
Dicke model to discuss how polariton condensation emerges in the
absence of carrier condensation.  In terms of numerical results, we
have presented superfluid fractions, atom density fluctuations and
zero-momentum occupation, and analyzed the phase diagram under
variation of $\mu_1$.  In addition, we have addressed the effects of
finite cluster size and the photon cutoff. This helps illustrate
connections to work on Jaynes--Cummings--Hubbard models and coupled
cavity arrays.  This topic has broad connections to other problems of
current interest including atom--molecule mixtures and the BEC-BCS
crossover in bosonic systems.

There are many avenues for further research including non-equilibrium
aspects and collective excitations
\cite{Andreev:Noneqfesh,Barankov:Coll}.  It would be interesting
to develop numerical techniques to explore finite temperature
polariton condensation and the onset of phase coherence in the Mott
phase. It would also be worthwhile to examine the phase diagram with
softcore bosons with finite $U_{aa}$ and $U_{bb}$.

\begin{acknowledgments}
  We are grateful to G. Conduit, N. Cooper, J. Keeling, and M. K\"ohl
  for helpful discussions. MJB, AOS, and BDS acknowledge EPSRC grant
  no. EP/E018130/1.
\end{acknowledgments}

\appendix

\section{Dicke Models}
\label{App:Dicke}
Throughout this manuscript we make use of the Dicke model
\cite{Dicke:Coherence} and its reductions. In the literature this
appears in various guises and with different names so we present a
brief guide.  The Dicke model describes $N$ two-level systems or
``spins'' coupled to radiation
\begin{equation}
H=\omega_0\sum_i^N S_i^z+\omega \psi^\dagger \psi+\frac{\bar
g}{\sqrt{N}}\sum_i^N(\psi^\dagger S_i^-+S_i^+\psi),
\label{HDicke}
\end{equation}
where ${\bf S}_i$ is a spin-$1/2$ operator and $\psi$ is a canonical
boson. Here $\omega_0$ is the level-splitting between the two-level
systems, $\omega$ is the frequency of the cavity mode, and $\bar g$ is
the strength of the matter--light coupling.  In quantum optics the
Dicke model is also known as the Tavis--Cummings model
\cite{Tavis:TC}. In the special case of a single spin, $N=1$, the
Dicke model is often referred to as the Jaynes--Cummings model
\cite{Jaynes:Comp}.  The Hamiltonian (\ref{HDicke}) is written in the
so-called rotating wave approximation in which terms of the form
$S_i^\pm\psi^\pm$ are excluded.  In the absence of these terms, the
model is integrable \cite{Hepp:Super,Hepp:Eqbm,Bog:TC} and exhibits a
superradiance quantum phase transition when $\bar
g=\sqrt{\omega\omega_0}$
\cite{Dicke:Coherence,Hepp:Super,Wang:Dicke,Hepp:Eqbm}; for a
discussion of the model including counter-rotating terms see
Ref.~\cite{Emary:Chaos}. Depending on the application these models are
often recast in different representations and we gather a few results
below.

\subsection{Jaynes--Cummings}
The simplest case is the Dicke model with $N=1$, or the
Jaynes--Cummings model, describing a single two-level system coupled
to radiation. It plays a central role in the variational analysis
presented in section \ref{Sect:Zerovar}, and recent problems in
matter--light systems, e.g. \cite{Greentree:QPTlight,Koch:JCL}.  A
convenient representation is the single site Hamiltonian
\begin{equation}
H=\epsilon_a n_a+\epsilon_b n_b+g(b^\dagger a +a^\dagger b),
\end{equation}
where $a$ is a hardcore boson (restricted to occupancy zero or one)
and $b$ may take arbitrary occupancy. We may construct exact
eigenstates of total integer occupation, $n=n^a+n^b\ge 1$, as
superpositions
\begin{equation}
|n\rangle=\alpha |1,n-1\rangle+\beta|0,n\rangle,
\end{equation}
where the first and second entries of the states are the occupations
of the $a$ and $b$ particles respectively. In matrix form this yields
the eigenvalue problem
\begin{equation}
\begin{pmatrix} \epsilon_a+(n-1)\epsilon_b & g\sqrt{n} \\ g\sqrt{n} & 
n\epsilon_b\end{pmatrix}\begin{pmatrix} \alpha \\ \beta \end{pmatrix}=
E_n\begin{pmatrix} \alpha \\ \beta \end{pmatrix},
\end{equation}
where we recall that $b|n\rangle=\sqrt{n}|n-1\rangle$ and
$b^\dagger|n\rangle=\sqrt{n+1}|n+1\rangle$ for bosons. The eigenvalues
are given by
\begin{equation}
E_n^\pm=n\epsilon_b-\omega_0/2\pm \sqrt{\omega_0^2/4+g^2n},
\label{JCeval}
\end{equation}
where $\omega_0\equiv \epsilon_b-\epsilon_a$. This agrees with the
spectrum of the Jaynes--Cummings model given by Greentree {\em et al}
\cite{Greentree:QPTlight}, where we identify $\epsilon_a$,
$\epsilon_b$, and $g$ with their parameters $\epsilon$, $\omega$, and
$\beta$. The coefficients in the corresponding normalized eigenvectors
$|n\rangle_\pm$ may be written
\begin{equation}
\begin{aligned}
\alpha_\pm & =
\frac{-\omega_0/2\pm\chi(n)}{\sqrt{2\chi^2(n)\mp\omega_0\chi(n)}},\\
\beta_\pm & = \frac{g\sqrt{n}}{\sqrt{2\chi^2(n)\mp\omega_0\chi(n)}},
\label{alphabeta}
\end{aligned}
\end{equation}
where $\chi(n)\equiv \sqrt{\omega_0^2/4+g^2n}$. We note that there is
a minor typing error in the eigenstates given in the methods section
of Ref.~\cite{Greentree:QPTlight}, where the basis states
$|1,n-1\rangle$ and $|0,n\rangle$ are erroneously
reversed. Equivalently, one may reverse the sign of
$\Delta\equiv\omega_0$ in their expression.

\subsection{Dicke Model}
\begin{figure}
\begin{center}
\includegraphics[width=8.2cm]{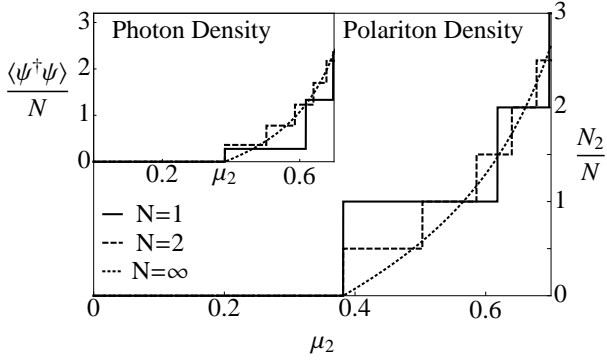}
\end{center}
\caption{Evolution of the polariton density, $N_2/N$, in the Dicke
model for $N=1,2$ sites and comparison to the $N\rightarrow \infty$
thermodynamic limit.  For $N=1$ the Dicke model reduces to the
Jaynes--Cummings model and the quantized polariton steps correspond to
the Mott lobes discussed in Ref.~\cite{Greentree:QPTlight}. In the
Dicke model these steps are quantized in units of $1/N$ and as
$N\rightarrow\infty$ we approach the variational results. Inset:
Evolution of the photon density. The magnetization (or population
imbalance) may be obtained by subtraction. All figures show the onset
of superradiance at $\mu_2=(3-\sqrt{5})/2\approx 0.382$, for
$\omega\rightarrow \tilde\omega=1-\mu_2$, $\omega_0\rightarrow
\tilde\omega_0 = 2-\mu_2$, and $\bar g=1$.}
\label{Fig:FiniteDicke}
\end{figure}

The generic Dicke model (\ref{HDicke}) is integrable for arbitrary $N$
\cite{Hepp:Super,Hepp:Eqbm,Bog:TC}, but is most conveniently analyzed
in the thermodynamic limit, $N\rightarrow\infty$ \cite{Hillery:Semi}.
However, in numerical simulations we must deal with the effects of
finite $N$ and truncations of the photon Hilbert space.  In
Fig.~\ref{Fig:FiniteDicke} we show the evolution of the polariton
density, $N_2/N$, where $N_2=\psi^\dagger\psi+\sum_i(S_i^z+1/2)$, and
the photon density as a function of $N$. This is obtained by solution
of the finite dimensional matrix problem. Both plots exhibit discrete
jumps which track the thermodynamic results. For $N=1$ the Dicke model
reduces to the Jaynes--Cummings model and the quantized polariton
steps correspond to the Mott lobes discussed in
Ref.~\cite{Greentree:QPTlight}.
 
\section{Zero Hopping Phase Diagram with $g=0$}
\label{Sect:Absence}
\begin{figure}
\begin{center}
\includegraphics[width=7.5cm,clip=true]{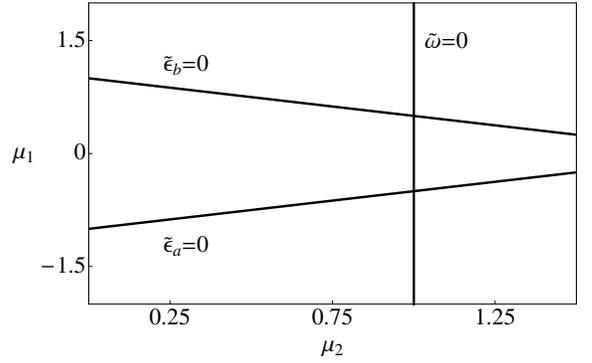}
\end{center}
\caption{Zero hopping phase diagram for the Hamiltonian
(\ref{reduced}) in the absence of matter--light interaction, with
$\epsilon_a=-1$, $\epsilon_b=1$, $\omega=1$, $g=0$. The loci
$\tilde\epsilon_a=0$, $\tilde\epsilon_b=0$, $\tilde\omega=0$,
indicated by the lower, upper and vertical lines respectively,
delineate the regions of population onset for $a,b,\psi$ in the grand
canonical ensemble.  When $g$ is switched on the horizontal boundaries
bend downwards and evolve into those shown in
Fig.~\ref{Fig:zerohop}. \label{Fig:Loci}}
\end{figure}

An instructive way to think about the topology of the zero hopping
phase diagram in Fig.~\ref{Fig:zerohop}, is in the absence of the
matter--light coupling. In Fig.~\ref{Fig:Loci} we plot the loci
$\tilde\epsilon_a=0$, $\tilde\epsilon_b=0$, $\tilde\omega=0$,
corresponding to population transitions in the Hamiltonian
(\ref{reduced}) for $g=0$. When $g$ is switched on, the horizontal
boundaries bend downwards and continuously evolve into those shown in
Fig.~\ref{Fig:zerohop}.

\section{Variational Phase Boundaries}
\label{App:varpb}
The variational energy for the finite hopping
problem allows some analytic progress with the phase boundaries, and
highlights connections to the Dicke model.  In the absence of
competition from other phases, the transition between the
non-superradiant insulator ($\theta=\chi=\bar\gamma=0$) and the
$a$-type superfluid ($\theta\neq 0$, $\chi=\eta=\bar\gamma=0$) for
example occurs when $\tilde\epsilon_a+zJ=0$. This may be seen by
explicit computation of the energies of each phase.  The energy
density in the generic Mott phase is
\begin{equation}
{\mathcal E}_{\theta=0}=(\tilde\epsilon_+-\tilde\epsilon_-\cos 2\chi)-
\frac{\bar g^2\sin^2 2\chi}{4\tilde\omega}.
\end{equation}
Minimizing on $\chi$ yields either $\chi=0$, corresponding to an
ordinary Mott insulator, with ${\mathcal E}_{\rm
MI}=\tilde\epsilon_+-\tilde\epsilon_-$, or
\begin{equation}
\cos 2\chi=\frac{2\tilde\omega\tilde\epsilon_-}{\bar g^2}\equiv
\frac{\tilde\omega\tilde\omega_0}{\bar g^2},
\end{equation}
corresponding to the superradiant Mott insulator. At this non-trivial
stationary point
\begin{equation}
  \frac{\partial^2\mathcal E}{\partial\chi^2}=\frac{2\left(\bar g^4-\tilde\omega^2\tilde\omega_0^2\right)}{\tilde\omega\bar g^2}.
\end{equation}
This corresponds to a minimum (with $\tilde\omega>0$) provided $\bar
g>\sqrt{\tilde\omega\tilde\omega_0}$. This coincides with the
superradiance transition in the Dicke model
\cite{Dicke:Coherence,Hepp:Super,Wang:Dicke,Hepp:Eqbm}. The energy
density in the superradiant Mott phase is
\begin{equation}
{\mathcal E}_{\rm SRMI}={\mathcal E}_{\rm MI}-\frac{\left(\bar g^2-\tilde\omega\tilde\omega_0\right)^2}{4\bar g^2\tilde\omega}.
\end{equation}
Similarly, in the $a$-type superfluid
\begin{equation}
{\mathcal E}_{\rm aSF}=(\tilde\epsilon_+-\tilde\epsilon_-)\cos^2\theta
-\frac{zJ}{4}\sin^22\theta.
\end{equation}
Minimizing on $\theta$ yields either $\theta=0$ or 
$\cos 2\theta=-\frac{(\tilde\epsilon_+-\tilde\epsilon_-)}{zJ}$.
The latter yields 
\begin{equation}
{\mathcal E}_{\rm aSF}=-\frac{(\tilde\epsilon_+-\tilde\epsilon_- -zJ)^2}{4zJ}
\end{equation}
The condition ${\mathcal E}_{\rm MI}={\mathcal E}_{\rm aSF}$ yields
the MI--aSF boundary $\tilde\epsilon_+-\tilde\epsilon_-+zJ=0$ or
equivalently $\tilde\epsilon_a+zJ=0$.  This meets the superradiance
onset $\bar g^2=\tilde\omega\tilde\omega_0$ at $\bar
g^2=\tilde\omega(\tilde\epsilon_b+zJ)$ or
\begin{equation}
\mu_1=\epsilon^b-\frac{\mu_2}{2}-\frac{\bar g^2}{\omega-\mu_2}+zJ.
\end{equation}
In a similar fashion, the transition between the superradiant Mott
state and the a-type superfluid occurs when ${\mathcal E}_{\rm
SRMI}={\mathcal E}_{\rm aSF}$ or
\begin{equation}
zJ(\bar g^2-\tilde\omega\tilde\omega_0)^2
=\bar g^2\tilde\omega(\tilde\epsilon_a+zJ)^2.
\label{SRMIASF}
\end{equation}

\section{Absence of Higher Lobes in Coupling to Quantum Light}
\label{App:Absence}
In Fig.~\ref{Fig:zerohop} we see that only the Mott lobe with atomic
density, $n=1$, is supported due to the coupling to the quantum light
field. To prove that only this lobe survives we compare the energies
of the higher states to those depicted in
Fig.~\ref{Fig:zerohop}. Without loss of generality we may focus on the
stable region defined by
$\tilde{\epsilon}_b-\bar{g}^2/4\tilde{\omega}>0$. Within this domain
$E^-_n(\gamma_{\rm var})$ and $E^-_n(\gamma=0)$ are both smoothly
increasing functions of $n$. Furthermore, whilst the super-radiance
condition (\ref{gc}) holds, $E^-_n(0)\geq E^-_n(\gamma_{\rm var})$. We
immediately conclude that whenever the $n=1$ superradiance condition
is met ($\bar{g}^4>\tilde{\omega}^2\tilde{\omega}^2_0$), the ground
state energy is either $E^-_1(\gamma_{\rm var})$ or $0$, the vacuum
energy. However, outside of this region we must examine whether the
lowest energy superradiant state is lower in energy than either
$E^-_1(0)$ or the vacuum.

To address this let us consider superradiant states with $n>1$. These
are candidates for the ground state within the region,
$\bar{g}^4<\tilde{\omega}^2\tilde{\omega}^2_0<n^2\bar{g}^4$, where the
upper bound follows from the superradiance condition and the lower
bound precludes the region in which $E^-_1(\gamma_{\rm var})$ is
supported. If $\tilde{\omega}_0>0$, taking roots of the inequality
yields $\tilde{\omega}_0>\tilde{\omega}\tilde{\omega}^2_0/n\bar{g}^2$
and $\tilde{\omega}_0>g^2/\tilde{\omega}$. In conjunction with the
condition for stability, we see that $E^-_{n}(\gamma_{\rm
var})>\tilde{\epsilon}_a=E^-_1(0)$. That is to say, the ground state
never has $n>1$.

If $\tilde{\omega}_0\leq 0$ a stronger statement can be made. Invoking
the superradiance condition,
$\tilde{\omega}_0<-\tilde{\omega}\tilde{\omega}^2_0/n\bar{g}^2$, it
follows that $E^-_n(\gamma_{\rm var})>0$ for \emph{all} $n$.  Hence
the vacuum must be the ground state when $\tilde{\omega}_0\leq
0$. (The latter implies
$E^-_1(0)=\tilde{\epsilon}_a\geq\tilde{\epsilon}_b>0$, where the last
inequality has followed from the stability requirement). We see that a
sufficient condition for termination of the Mott lobe is
$\mu_2>\omega_0$. However, since $\mu_2<\omega$ in general, such
regimes are only stable for $\omega_0<\omega$; see
Fig.~\ref{Fig:zerohop}.

\end{document}